\documentclass[preprint,12pt,3p]{elsarticle}

\usepackage{amsmath,amssymb}
\usepackage{graphicx}
\usepackage{color}
\usepackage{subcaption}
\usepackage{siunitx}
\usepackage{dcolumn}
\usepackage{bm}
\usepackage{pdfpages}

\usepackage{lineno}
\usepackage[usestackEOL]{stackengine}

\begin{document}

\begin{frontmatter}
\title{Computation of drug solvation free energy in supercritical CO$_2$: alternatives to all-atom computer simulations}

\author[isc]{Kalikin N.N. }
\author[isc,hse]{Budkov Y.A. \corref{cor1}}
\author[lpzg]{Kolesnikov A.L.}
\author[isc]{Ivlev D.V. }
\author[isc]{Krestyaninov M.A.}
\author[isc]{Kiselev M.G.}
\address[isc]{G.A. Krestov Institute of Solution Chemistry of the Russian Academy of Sciences, Laboratory of NMR Spectroscopy and Numerical Investigations of Liquids, Akademicheskaya Str. 1, 153045, Ivanovo, Russia}
\address[hse]{Tikhonov Moscow Institute of Electronics and Mathematics, School of Applied Mathematics, National Research University Higher School of Economics, 34, Tallinskaya Ulitsa, 123458, Moscow, Russia}
\address[lpzg]{Institut für Nichtklassische Chemie e.V., Permoserstr. 15, 04318, Leipzig, Germany}
\cortext[cor1]{Corresponding author}

\begin{abstract}
Despite the modern level of development of computational chemistry methods and technological progress, fast and accurate determination of solvation free energy remains a huge problem for physical chemists. In this paper, we describe two computational schemes that can potentially solve this problem. We consider systems of poorly soluble drug compounds in supercritical carbon dioxide. Considering that the biggest contribution among all intermolecular interactions is made by van der Waals interactions, we model solute and solvent particles as coarse-grained ones interacting via the effective Lennard-Jones potential. The first proposed approach is based on the classical density functional theory and the second one relies on molecular dynamics simulation of the Lennard-Jones fluid. Sacrificing the precision of the molecular structure description while capturing the phase behavior of the fluid with sufficient accuracy, we propose computationally advantageous paths to obtaining the solvation free energy values with the accuracy satisfactory for engineering applications. The agreement reached between the results of such coarse-graining models and the experimental data indicates that the use of the all-atom molecular dynamic simulations for the studied systems seems to be excessive. 
\end{abstract}

\end{frontmatter}

\section*{\sffamily \Large INTRODUCTION} 

Solvation free energy is an extremely important thermodynamic quantity for a wide range of physico-chemical applications, including calculation of solubility values, estimation of partition coefficients, infinite dilution activity coefficients and other parameters. Nowadays there are a number of computational methods developed to determine solvation free energy values. Regarding the systems that include poorly soluble drugs and drug-like compounds it is rather popular among scientists to rely on the results of the all-atom molecular dynamics (AAMD) simulations based on alchemical free energy methods \cite{noroozi2016solvation, frolov2015accurate, bruckner2011efficiency,hansen2014practical,jia2016calculations, misin2016hydration}. Such approaches model a set of nonphysical intermediate states to compute the values of free energy of solute molecule transfer from the gas phase to the solution. More details regarding different alchemical methods can be found elsewhere \cite{shirts2012best}. Despite their evident efficiency and similarly clear time expenditures and computational cost, these methods can have other drawbacks. They include high sensitivity to the force field choice and uncertainty arising from the choice of the partial atomic charges computation method. Both of these aspects of the AAMD calculation can lead to the inconsistency of the final solvation free energy values up to several kcal/mol\cite{lundborg2015automatic,noroozi2016solvation,jambeck2013partial}. At the same time, the procedure of the force-field parametrization is itself a nontrivial task and it is not always easy to justify some of the approaches physically. Thus, if the aim is to determine the free energy quickly, the utilization of the AAMD techniques is rather questionable. On the other hand, as it has been recently shown \cite{da2020all,glova2019toward,papavasileiou2019molecular,ewen2016comparison}, the results of the coarse-graining models using united-atom force fields are comparable in accuracy with those of the all-atom ones and in some cases even outperform them. Another approach to the estimation of drug solvation free energy in supercritical CO$_2$ (scCO$_2$), popular among chemical engineers, is based on applying cubic equations of state, for instance, the Peng-Robinson equation of state (PR EOS) for binary mixtures, which is a generalization of the classical van der Waals equation of state \cite{garlapati2009temperature,moine2019can}. However, the PR EOS contains several adjustable parameters, the determination of which requires an experimental data set (solubility isotherms, for instance). That is why the PR EOS is usually used to approximate the existing experimental data. Thus, in order to apply the PR EOS for solvation free energy estimation, it is necessary to perform preliminary experimental measurements that make this approach much more complicated. The next step toward a more robust description of the phase equilibrium is a family of SAFT (Statistical Associating Fluid Theory) type models \cite{kontogeorgis2020equations}. As they include the contributions of different interaction terms and the parameters of these models hold a strict physical meaning, they have been successfully utilized to correlate and predict the solvation properties of a variety of solutes  \cite{hutacharoen2017predicting,el2013application,anvari2014study,yang2005modeling,sodeifian2020experimental,mahmoudabadi2021application}. Although there is a rather substantial decrease in the number of binary interaction parameters, when multicomponent systems are considered, as compared to the cubic EOS with advanced mixing rules, the whole parametrization procedure is still cumbersome and the choice of the best one is itself a nontrivial task \cite{ramirez2020parameterization}. High accuracy of the calculated solvation free energy values with a error less than 1 kcal/mol for sufficiently large data sets can be also achieved by using quantum mechanical calculations on continuum solvents (COSMO-based models, SMD)\cite{shimoyama2009development,wang2014predictive,marenich2009universal,chamberlin2008modeling,klamt2005cosmo}. The cost of such accuracy, however, is the complex parametrization routine and the necessity of its replication for new systems, if they are dramatically different from the original learning set \cite{misin2016predicting}.

Quite promising is the method of solvation free energy estimation based on the classical density functional theory (cDFT). In papers \cite{sokolov2007fundamental,chuev2006hydration}, the authors proposed a method for calculating hydration free energy of a set of hydrophobic solutes. This method can be considered as a coarse-grained one, since within its framework real molecules of both the solute and the solvent are replaced by spherically symmetric particles interacting with each other through the effective Lennard-Jones (LJ) potential. The utilization of the Weeks-Chandler-Andersen (WCA) procedure for the LJ potential and application of the fundamental measure theory \cite{rosenfeld1989free} (FMT) to account for the short-range hard core correlations allowed the authors to fit the pressure and the surface tension of bulk water to the experimental values by changing the cutoff radius. The latter, in turn, enabled them to describe hydration of a set of hydrocarbons. Even though such cDFT approach remains a method of experimental data approximation, it has distinct advantages over the methods based on the implications of the PR EOS. The main one is that, in contrast to the PR EOS-based approach \cite{peng1976new}, which in its nature is a variant of the mean-field theory taking into account only long-range correlations of solution molecules, the cDFT-based method accounts for the short-range correlations of the solvent molecules due to the packing effects in the solute molecule solvation shell. Moreover, within the cDFT approach, all the interaction parameters are related directly to the effective LJ interaction potentials and, thereby, have a clear physical meaning.

The development, in a sense, of the cDFT application for the solvation free energy calculation can be traced by a number of papers \cite{zhao2011molecular,zhao2011new,sergiievskyi2014fast,gendre2009classical}, proposing the so-called molecular density functional theory approach, which consists of several steps. The first one is computation of the site-site pair-distribution functions from the AAMD simulation of the pure solvent. The second one is obtaining of the direct correlation function from the Molecular Ornstein–Zernike integral equation, and, finally, calculation of the energy functional, the value of which, being the difference between the system grand thermodynamic potential with and without the solute molecule, gives the solvation free energy of the solute. Although such approach is faster than the full-scale classical AAMD simulation, the first step is still dependent on the force field choice. The results obtained for a series of halide ions are in good agreement with the experiment, but the question is what results could be achieved for drug-like molecules and how the calculation of such systems would influence the computational speed.

In this paper, we discuss two coarse-graining techniques that can be used for fast and sufficiently accurate estimation of solvation free energy values of sparingly soluble drug compounds in a scCO$_2$ medium. One of the potential applications of the obtained data can be subsequent calculation of the solubility of the compounds, which is a prerequisite for using supercritical technologies, e.g. micronization or cocrystallization \cite{baghbanbashi2020solubility, padrela2018supercritical}, in order to enhance the final aqueous solubility and thus the bioavailability of drugs.     
The proposed approaches can make solvation free energy computation faster than in the aforementioned techniques, preserving the accuracy of the obtained values. 
The first one is based on cDFT, but, in contrast to the approaches discussed above, the only input values required to compute the interaction potential parameters are the critical parameters of the solvent and solute, which can be often found in literature. The second approach is based on the coarse-grained MD (CGMD) simulations of the LJ fluid, with the interaction parameters determined on the basis on the law of the corresponding states, knowing the critical parameters of the system compounds and critical parameters of the LJ fluid.  

\section*{\sffamily \Large METHODOLOGY}

\section*{\sffamily \Large Classical Density Functional Theory}

The details of the proposed density functional theory-based method are presented in Appendix. Here we briefly outline the main points. Within such approach CO$_2$ molecules are coarse-grained to spherically symmetric particles interacting through the effective pairwise LJ potential, the parameters of which can be obtained by fitting the CO$_2$ liquid-vapor critical point. Starting from the expression of the grand thermodynamic potential for the CO$_2$ fluid in the external potential field $V_{ext}(\mathbf r)$ created by the fixed solute molecule located at the origin, one gets
\begin{linenomath}
\begin{equation}\label{3rd_eq}
  \Omega[\rho(\mathbf r)]=F_{int}[\rho(\mathbf r)]+\int d\mathbf r \rho(\mathbf r)(V_{ext}(\mathbf r)-\mu),
\end{equation}
\end{linenomath}
where $F_{int}[\rho(\mathbf r)]$ is the intrinsic Helmholtz free energy of the CO$_2$ fluid and $\mu$ is the chemical potential of the bulk fluid at certain temperature and pressure values.
The intrinsic Helmholtz free energy, in its turn, can be comprised of two contributions 
\begin{linenomath}
\begin{equation}\label{4th_eq}
  F_{int}[\rho(\mathbf r)]=k_BT\int d\mathbf r\rho(\mathbf r)[\ln(\Lambda^3\rho(\mathbf r))-1]+F_{ex}[\rho(\mathbf r)],
\end{equation}
\end{linenomath}
where the first one is the Helmholtz free energy of the ideal gas and the second one is the excess Helmholtz free energy of the fluid; $\Lambda$ is the thermal de Broglie wavelength.
One can write the total excess free energy in the form
\begin{equation}
\label{F_ex}
F_{ex}[\rho(\mathbf r)]=F_{hs}[\rho(\mathbf r)]+F_{att}[\rho(\mathbf r)].
\end{equation}
The hard spheres contribution (the first term on the right hand side) is determined within Rosenfeld's version of the fundamental measure theory (FMT) \cite{rosenfeld1989free} as follows
\begin{equation}\label{6th_eq}
F_{hs}[\rho(\mathbf r)]=k_BT\int d\mathbf r\Phi(\mathbf r),
\end{equation}
where $\Phi(\mathbf{r})$ is the excess free energy density (expressed in $k_B T$ units), which is the function of weighted densities. The attractive contribution to the excess free energy is described within the mean-field approximation \cite{archer2017standard}
\begin{equation}\label{11th_eq}
  F_{att}[\rho(\mathbf{r})]=\frac{1}{2}\int d\mathbf{r}\rho(\mathbf{r})\int d\mathbf{r}'\rho(\mathbf{r}')\phi_{WCA}(\mathbf{r}-\mathbf{r}'),
\end{equation}
with the effective WCA pair potential of the attractive interactions.

The solute molecule is modeled as a particle creating an external LJ potential with the effective parameters of the interaction between this molecule and the molecules of the fluid, which are determined according to the Berthelot-Lorenz mixing rules: $\sigma_{sf}=(\sigma_{ss}+\sigma_{ff})/2$ and $\varepsilon_{sf}=\sqrt{\varepsilon_{ss}\varepsilon_{ff}}$.  
The total pressure of the fluid bulk phase takes the following form
\begin{linenomath}
\begin{equation}
    P=\rho_{b} k_BT\frac{1+\eta+\eta^2}{(1-\eta)^3}+\frac{1}{2}B_{WCA}\rho_b^2.
\end{equation}
\end{linenomath}
where $\eta=\pi d_{BH}^3\rho_b/6$ is the packing fraction of the hard sphere system, the effective Barker-Henderson (BH) diameter is determined by the Pade approximation \cite{Verlet1972a}:$d_{BH} = \sigma_{ff}(1.068 \varepsilon_{ff}/k_B T + 0.3837)/(\varepsilon_{ff}/k_B T + 0.4293)$, and the following auxiliary function corresponding to the attractive contribution is introduced
\begin{linenomath}
\begin{equation}
    B_{WCA}=-\frac{32\sqrt2}{9}\pi\varepsilon_{ff}\sigma_{ff}^3+\frac{16}{3}\pi\varepsilon_{ff}\sigma_{ff}^3\left[\left(\frac{\sigma_{ff}}{r_c}\right)^3-\frac{1}{3}\left(\frac{\sigma_{ff}}{r_c}\right)^9\right].
\end{equation}
\end{linenomath}
The parameters of the interaction potential between two molecules of CO$_2$ ($\sigma_{ff}$, $\varepsilon_{ff}$) can be obtained, as it was mentioned above, by fitting the respective parameters of the liquid-gas critical point, i.e. solving the following system of equations, using the known critical parameters of CO$_2$
\begin{linenomath}
\begin{equation}
    \frac{\partial P}{\partial\rho_b}\Bigr|_{\substack{T=T_c\\\rho_b=\rho_c}}=0, ~~ \frac{\partial^2 P}{\partial\rho_b^2}\Bigr|_{\substack{T=T_c\\\rho_b=\rho_c}}=0.
\end{equation}
\end{linenomath}
In the same manner one can find the parameters of the LJ potential for the active compound ($\sigma_{ss}$, $\varepsilon_{ss}$). NIST \cite{nist} provides accurate values of the critical parameters for CO$_2$. At the same time, the critical parameters of the solutes can be only estimated. The values of the calculated potential parameters are presented in Table \ref{table01}. The problems that can potentially arise from the ambiguities of critical parameter estimation are discussed in the next sections.
After the iterative minimization of the grand thermodynamic potential with respect to the density and finding its equilibrium profile, one can obtain the solvation free energy as the excess grand thermodynamic potential:
\begin{equation}
  \Delta G_{solv} = \Omega[\rho(\mathbf r)] - \Omega[\rho_b].
\end{equation}
We would like to note that $\Omega[\rho_b]$ is calculated at $V_{ext}(\mathbf{r})=0$.

\begin{table}[h]
\centering
\caption{Parameters of the LJ potential of CO$_2$ and drug compounds obtained from the fitting of the corresponding critical points using EOS implemented within the cDFT-based approach.}
\begin{tabular}{c|c|c}
compound  & $\sigma$, $[\si{\angstrom}]$ & $\varepsilon$, [K]  \\
\hline
CO2             &   3.363                                        & 218.738    \\
aspirin         &   6.553                                        & 548.700    \\
ibuprofen       &   7.301                                        & 539.206    \\ 
carbamazepine   &   7.180                                        & 565.911    \\
\end{tabular}
\label{table01}
\end{table}

\section*{\sffamily \Large All-atom MD simulation}

For comprehensive investigation we have conducted a set of AAMD simulations for the system of aspirin (ASP) in scCO$_2$ to find out how the choice of the AAMD simulation parameters affects the final solvation free energy values. We have compared the results for three different force fields describing the ASP molecule – GROMOS \cite{malde2011automated,stroet2018automated}, GAFF \cite{wang2004development} and OPLS-AA \cite{kaminski2001evaluation}, and two models describing the CO$_2$ molecules -  Zhang's \cite{zhang2005optimized} and TraPPE \cite{potoff2001vapor} ones. We have established that the absolute values of the ASP solvation free energy obtained using the GROMOS force field with the help of the Automated Topology Builder system \cite{malde2011automated,canzar2013charge,koziara2014testing} are abnormally high when compared to those calculated by the two other force-fields. As the procedure of accurate force field parametrization for obtaining precise final values was not the goal of the present study, rather, we wanted to test the performance of the generic force fields without preliminary preparations, the results obtained with the GROMOS force field will not be discussed. 

The simulation details are similar to those we used in our previous studies \cite{budkov2019possibility,kalikin2020carbamazepine}. Here we outline the main aspects. The AAMD simulations were conducted in the Gromacs 4.6.7 program package \cite{pronk2013gromacs,abraham2015gromacs,bekker1993gromacs,berendsen1995gromacs} $\textrm{[http://www.gromacs.org]}$. The cell contained 1024 CO$_2$ molecules and 1 ASP molecule. The cell was constructed in the Packmol program \cite{martinez2009packmol}, the CO$_2$ density at different state parameters was taken from the NIST database \cite{nist}. The Bennett acceptance ratio (BAR) \cite{bennett1976efficient} approach was chosen to compute the free energy surface. For each simulation we performed equilibration for 100 ps with a 1 fs step and 500 ps with a 2 fs step in the NVT and NPT ensembles, respectively, and the production run of 5 ns with the step of 2 fs.

We calculated partial atomic charges for the ASP molecule by the Merz-Kollman method \cite{singh1984approach,besler1990atomic} using Gaussian 09 software \cite{frisch2013gaussian} with the PBE functional and $\mathrm{6-311+}$ $\mathrm{+g(2d,p)}$ basis set. We averaged partial atomic charges over two most stable ASP conformers (see Appendix). Such procedure improved the agreement between the solvation free energy values of carbamazepine (CBZ) obtained experimentally and computed within the AAMD simulation in our previous paper \cite{kalikin2020carbamazepine}.

For each computation of the solvation free energy surface we conducted 12 independent simulations, each corresponding to a different pair of coupling parameters, describing the interactions between the solvent molecules and the solute molecule. The potential function is linearly dependent on the coupling parameters. The LJ solute-solvent interaction parameters were calculated from the atomic parameters by the Berthelot-Lorentz mixing rules. We have chosen the following set of alchemical coefficients $\{\lambda_{LJ},\lambda_{C}\}$: $\{0.0,0.0\}$, $\{0.2,0.0\}$, $\{0.5,0.0\}$, $\{1.0,0.0\}$, $\{1.0,0.2\}$, $\{1.0,0.3\}$, $\{1.0,0.4\}$, $\{1.0,0.5\}$, $\{1.0,0.6\}$, $\{1.0,0.7\}$, $\{1.0,0.8\}$, $\{1.0,1.0\}$, where the subscripts "C" and "LJ" correspond to the scaling parameters of the Coulomb and Lennard-Jones potentials. The differences in the free energy between the states with neighboring values of the coupling parameters were calculated using the “g$\_$bar” tool of the Gromacs package.

\section*{\sffamily \Large Coarse-grained MD simulation}

Besides the cDFT-based approach discussed above, we also propose a method of total coarse-graining of the molecular structure of the sparingly soluble drug compounds for accelerated computation of the solvation free energy using MD simulations. It represents a mixture of the fluid and solute molecules as a LJ fluid. The interaction parameters are determined by the law of the corresponding states. Thus, knowing the critical parameters of the solvent and solute, we can fit the critical point of the LJ fluid, obtaining the parameters of the potential of the solute-solute and solvent-solvent interactions. The parameters of the solute-solvent interaction potential are determined by the standard Berthelot-Lorenz mixing rules. The reduced quantities can be obtained in the standard way as follows:
$T^{\ast}=T/\varepsilon$ - temperature, $\rho^{\ast}=\rho\sigma^3$ - density and $P^{\ast}=P\sigma^3/\varepsilon$ - pressure. The reduced critical parameters for the pure and untruncated LJ fluid were calculated by Monte Carlo
simulations in the grand canonical ensemble \cite{potoff1998critical}: $T_c^*=1.3120$, $\rho_c^*=0.316$ and $P_c^*=0.1279$. The obtained parameters of the LJ potential for CO$_2$ and three studied compounds are presented in Table \ref{table02}.

The simulation procedure is the same as the one described in the previous section for AAMD, except for the change in the set of the scaling $\lambda$ parameters. As we consider only the LJ contribution for such simulations, the parameters take the following values $\{\lambda_{LJ},\lambda_{C}\}$:$\{0.0,0.0\}$, $\{0.1,0.0\}$, $\{0.2,0.0\}$, $\{0.3,0.0\}$, $\{0.4,0.0\}$, $\{0.5,0.0\}$, $\{0.6,0.0\}$, $\{0.7,0.0\}$, $\{0.8,0.0\}$, $\{0.9,0.0\}$, $\{1.0,0.0\}$.

\begin{table}[h]
\centering
\caption{Parameters of the LJ potential of the CO$_2$ and drug compounds used in the CGMD simulations and obtained from the fitting of the LJ fluid liquid-vapor critical point.}
\begin{tabular}{c|c|c}
compound  & $\sigma$, $[\si{\angstrom}]$ & $\varepsilon$, $[K]$  \\
\hline
CO2             &   3.673                                        & 231.805    \\
aspirin         &   6.789                                        & 581.479    \\
ibuprofen       &   7.565                                        & 571.387    \\ 
carbamazepine   &   7.439                                        & 599.718    \\
\end{tabular}
\label{table02}
\end{table}

\section*{\sffamily \Large Experimental solvation free energy}

To compare the calculated values with the experimental results we propose to extract the solvation free energy values from the experimentally measured solubility data available in literature. As the studied compounds are sparingly soluble in a supercritical fluid, we consider the limit of their infinite dilution. One should also assume that a supercritical fluid does not dissolve in the solid phase, the solute molar volume does not depend on pressure and the fugacity coefficient of the solid phase can be equal to unity due to the fact that the sublimation pressure of the solute is rather small. Thus, following from the condition of the chemical equilibrium for the solute between its solid and solution phases (the details of derivation and discussion can be found elsewhere \cite{noroozi2016solvation,baghbanbashi2020solubility,hartono2001prediction,de2009solid,su2006simulations}), one can obtain the solvation free energy in the form
\begin{equation}
    \Delta G_{solv}=RT\ln{\frac{P^{sat}}{y_2\rho_b RT}}+v^s(P-P^{sat}),
    \label{deltaG}
\end{equation}
where $\Delta G_{solv}$ is the solvation free energy of the compound in the scCO$_2$ medium, $R$ is the gas constant, $T$ is the temperature, $P^{sat}$ is the sublimation pressure of the solute, $y_2$ is the solubility of the compound expressed in molar fraction, $\rho_b$ is the bulk density of the fluid, $v^s$ is the molar volume of the solute and $P$ is the total pressure, imposed in the system. It can be concluded from expression (\ref{deltaG}) that apart from the experimental data on solubility, one has to know the properties of the pure compound, i.e. the sublimation pressure and the molar volume. The former can be estimated by semi-empirical methods (see, for example, \cite{komkoua2013evaluation,o2014assessment} and cites therein) or, which is more preferable, can be measured experimentally \cite{perlovich2004thermodynamics,perlovich2004solvation,drozd2017novel}, and the latter one is often estimated by the group contribution approaches (see \cite{cao2008use}). The sources of the experimental data discussed in the paper for all the solutes are presented in Table \ref{table10}.

\begin{figure}[h]
\center{\includegraphics[width=0.8\linewidth]{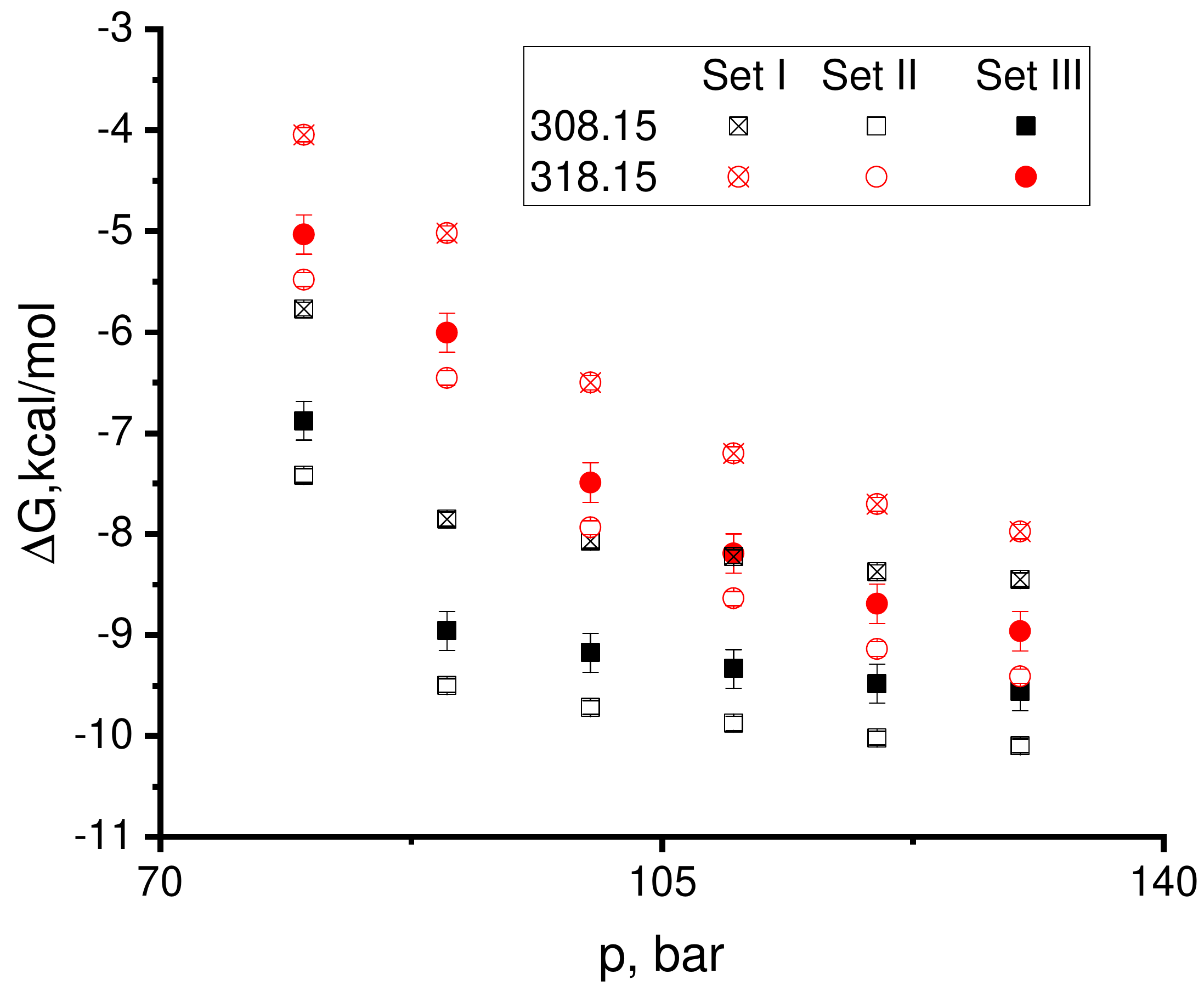}}
\caption{Comparison of the IBU solvation free energy values, extracted from the experiment \cite{ardjmand2014measurement}, using three sets of the sublimation pressure data from Table \ref{table1}.}
\label{fig_002}
\end{figure}

In Fig. \ref{fig_002} we show a comparison of the solvation free energy values obtained from the ibuprofen (IBU) solubility experiment \cite{ardjmand2014measurement}, using three sets of sublimation pressure values, which we will designate as set Ardjmand\_setI\cite{ardjmand2014measurement}, Ardjmand\_setII \cite{garlapati2009temperature} and Ardjmand\_setIII \cite{perlovich2004thermodynamics} (Table \ref{table1}). The error of the final solvation free energy measurements was calculated as the uncertainty of the indirect measurements, the main contribution to which turned out to mainly depend on the uncertainty of the sublimation pressure values. As it can be seen, the data are rather scattered and several values reach the discrepancies of up to 1.5 $kcal/mol$. The molar volume was estimated \cite{baum1997chemical} to be $182.1\times 10^{-6}$ $m^3/mol$. It should be noted that the values of set III were measured experimentally by the transpiration method \cite{zielenkiewicz1999vapour}, the values of set I were numerically approximated \cite{lyman1990handbook}, and in the case of set II the sublimation pressure was considered as an adjustable parameter, the data for which were obtained by minimizing the discrepancy between the experimental solubility data and those correlated by the equation of state. In the case when strict experimental data are available in literature, which is the case for all the compounds studied here, we prefer to rely on them. Thus, in the following sections we will utilize only the experimentally measured sublimation pressure values.

\begin{table}[h]
\centering
\caption{Experimental solvation free energy data.}
\begin{tabular}{c|c|c|c|c}
label   &  compound  & \Centerstack{ solubility data \\source} & \Centerstack{ sublimation pressure \\ source} & \Centerstack{ molar volume \\ source}   \\
\hline
Ardjmand\_setI & IBU &   \cite{ardjmand2014measurement}  &   \cite{ardjmand2014measurement}    &   \cite{ardjmand2014measurement} \\
Ardjmand\_setII & IBU &   \cite{ardjmand2014measurement}  &   \cite{garlapati2009temperature}    &     \cite{ardjmand2014measurement}\\
Ardjmand\_setIII &   IBU &   \cite{ardjmand2014measurement}  &   \cite{perlovich2004thermodynamics}  &   \cite{ardjmand2014measurement}\\
Charoenchaitrakool  &   IBU &   \cite{charoenchaitrakool2000micronization}  &   \cite{perlovich2004solvation}   &   \cite{ardjmand2014measurement}\\
Huang   &   ASP &   \cite{huang2004solubility}  &   \cite{perlovich2004solvation} &  \cite{huang2004solubility} \\
Champeau    &   ASP &   \cite{champeau2016solubility}   &   \cite{perlovich2004solvation}   &  \cite{huang2004solubility} \\
Yamini  &   CBZ &   \cite{yamini2001solubilities}   &   \cite{drozd2017novel}   &   \cite{li2013new} \\
IR  &   CBZ &   \cite{kalikin2020carbamazepine} &   \cite{drozd2017novel}   &   \cite{li2013new} \\
\end{tabular}
\label{table10}
\end{table}

\begin{table}[h]
\centering
\caption{Sublimation pressure values of the IBU compound extracted from literature.}
\begin{tabular}{c|c|c|c}
       & \multicolumn{3}{c}{$P^{sat}$, Pa}                        \\
\hline
T, K   &  Ardjmand\_setI \cite{ardjmand2014measurement}  & Ardjmand\_setII \cite{garlapati2009temperature} & Ardjmand\_setIII \cite{perlovich2004thermodynamics}   \\
\hline
308.15 &   0.0495                                           & 0.0033                                 & 0.0081 \\
313.15 &   0.0897                                           & 0.0075                                 & 0.0167 \\
318.15 &   0.1600                                           & 0.0165                                 & 0.0335 \\     
\end{tabular}
\label{table1}
\end{table}

\section*{\sffamily \Large Choice of the critical parameters}

\begin{figure}[h]
\center{\includegraphics[width=0.8\linewidth]{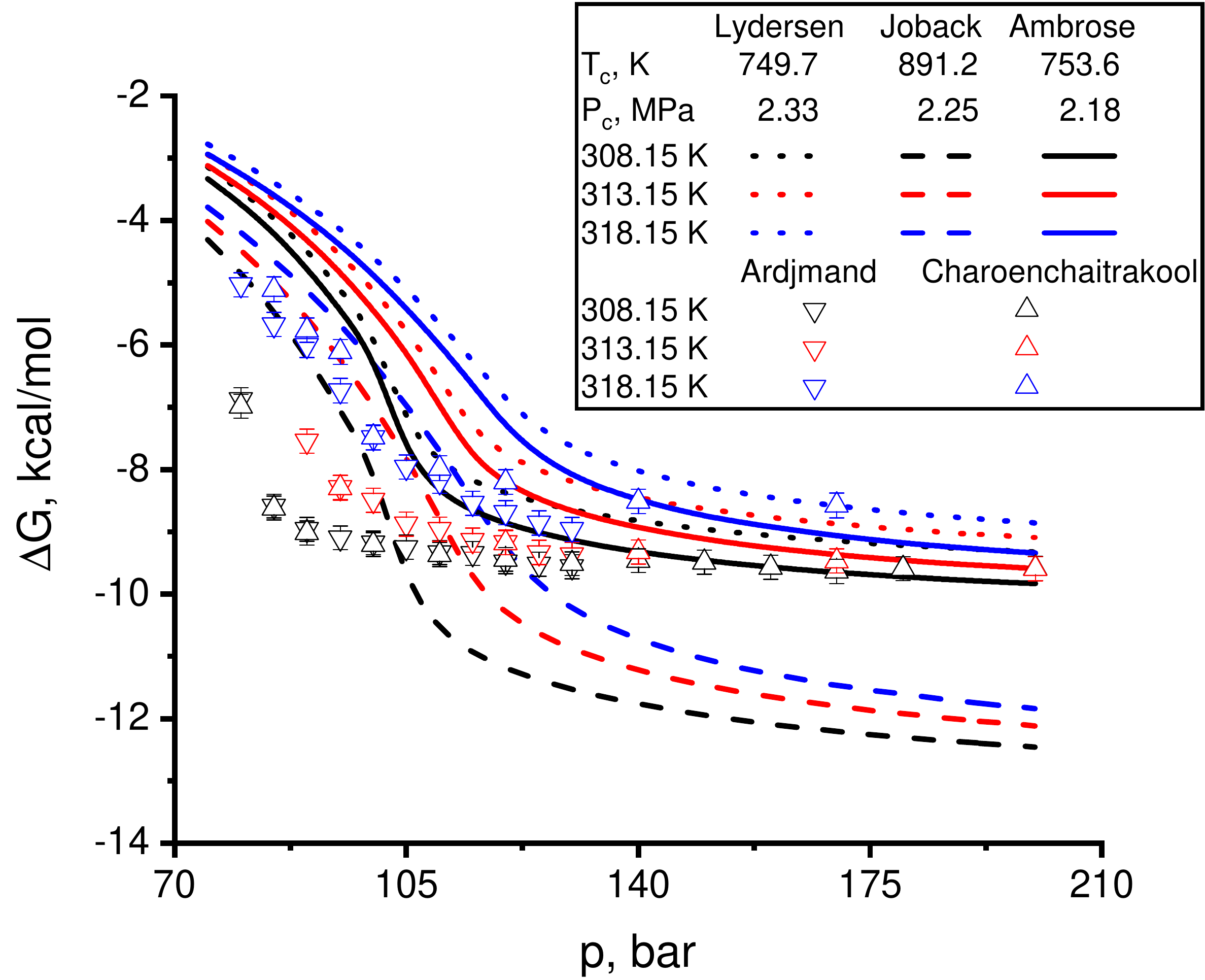}}
\caption{Comparison of the IBU solvation free energy values obtained by the cDFT approach using three sets of estimated critical parameters \cite{charoenchaitrakool2000micronization};  experimental data from \cite{ardjmand2014measurement,charoenchaitrakool2000micronization}.}
\label{fig_7}
\end{figure}

One of the advantages of the two proposed coarse-graining approaches is a small number of input parameters. In fact, one only needs to know the solute and solvent critical parameters for the system of choice. Finding them for a solvent is usually not difficult as accurate information for the most frequently used ones can be obtained from NIST \cite{nist}. The bottleneck of the approach is the fact that the critical parameters of a solute cannot be determined experimentally. It is common to estimate them by the group contribution methods extensively used for this purpose \cite{klincewicz1984estimation, joback1987estimation, constantinou1994new}. However, 
the values of the critical parameters are highly dependent on the calculation method. Thus, the final solvation free energy values can vary a lot, as we show in  Fig.\ref{fig_7}, where we compare the IBU solvation free energy values obtained by the cDFT approach for three isotherms using three sets of the estimated critical parameters \cite{charoenchaitrakool2000micronization}. The symbols on the plot corresponding to the experimental data suggest that the Ambrose method in this case gives the most adequate critical values. Nevertheless, it is obvious that the final values of the solvation free energy are highly dependent on the choice of the critical parameters and the change in the critical temperature seems to affect the outcomes more crucially than the change in the critical pressure. It has been recently shown \cite{jahromi2019estimation} that a possible solution to this ambiguity problem could be the method of the critical parameters estimation proposed by G.Kontogeorgis et al. \cite{kontogeorgis1997method} that more accurately predicts the sublimation enthalpy and vapor pressure than the methods developed by K.Joback \cite{joback1987estimation}, K.Klincewicz \cite{klincewicz1984estimation} and L.Constantinou \cite{constantinou1994new}. Moreover, the approach by Kontogeorgis is applicable in the cases of the compounds with pharmaceutical-like high molecular weights. However, while the latter three methods use the value of the boiling point, which in practice cannot be determined experimentally and thus has to be estimated as well, Kontogeorgis' method requires at least one experimental vapor pressure data point.                  

\section*{\sffamily \Large RESULTS AND DISCUSSION}

\section*{\sffamily \Large Coexistence curves}

\begin{figure}
 \begin{subfigure}[b]{0.8\linewidth}
    \centering\includegraphics[width=.8\linewidth]{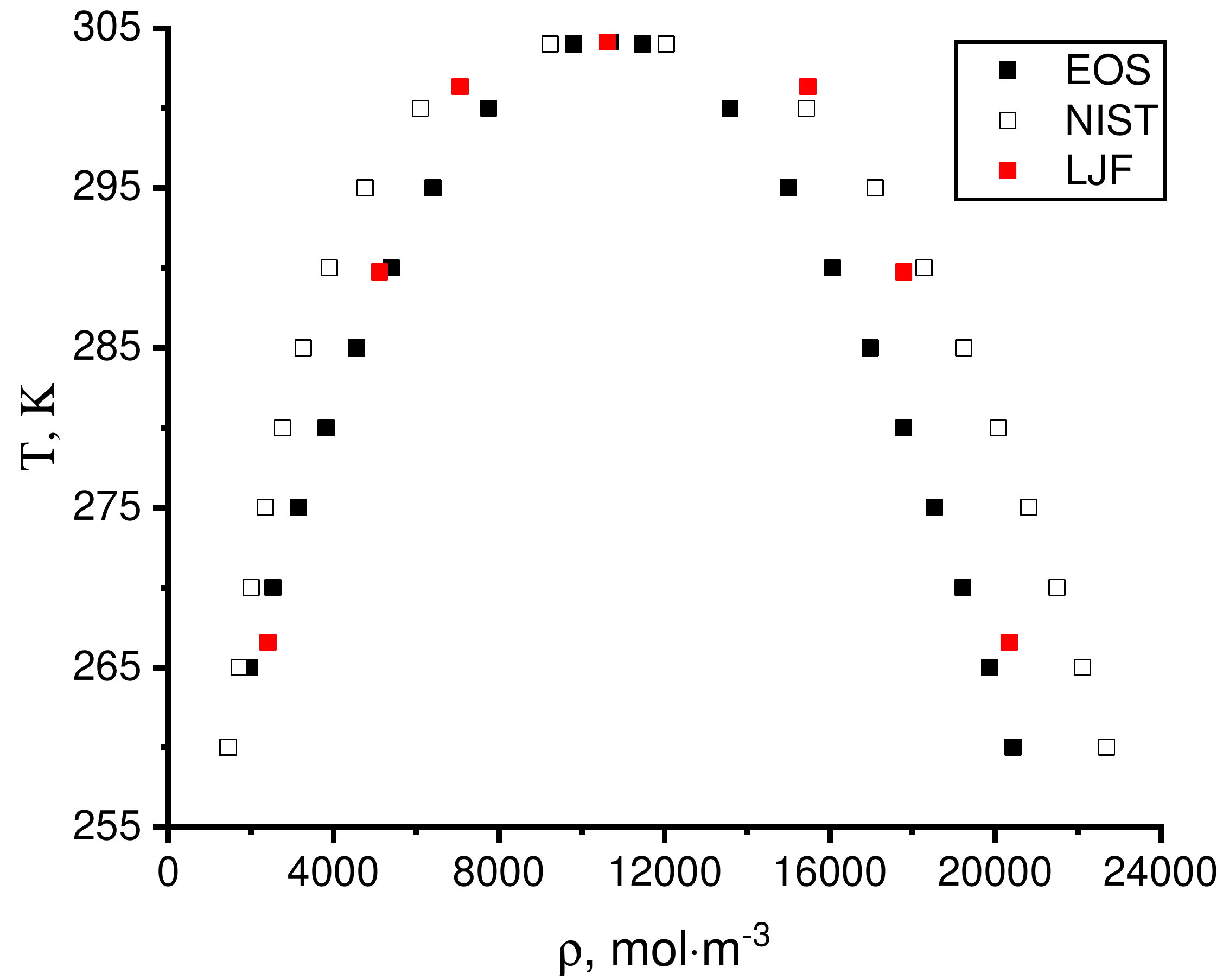}
    \caption{\label{fig01:fig1}}
  \end{subfigure}
  \newline
 \begin{subfigure}[b]{0.8\linewidth}
    \centering\includegraphics[width=.8\linewidth]{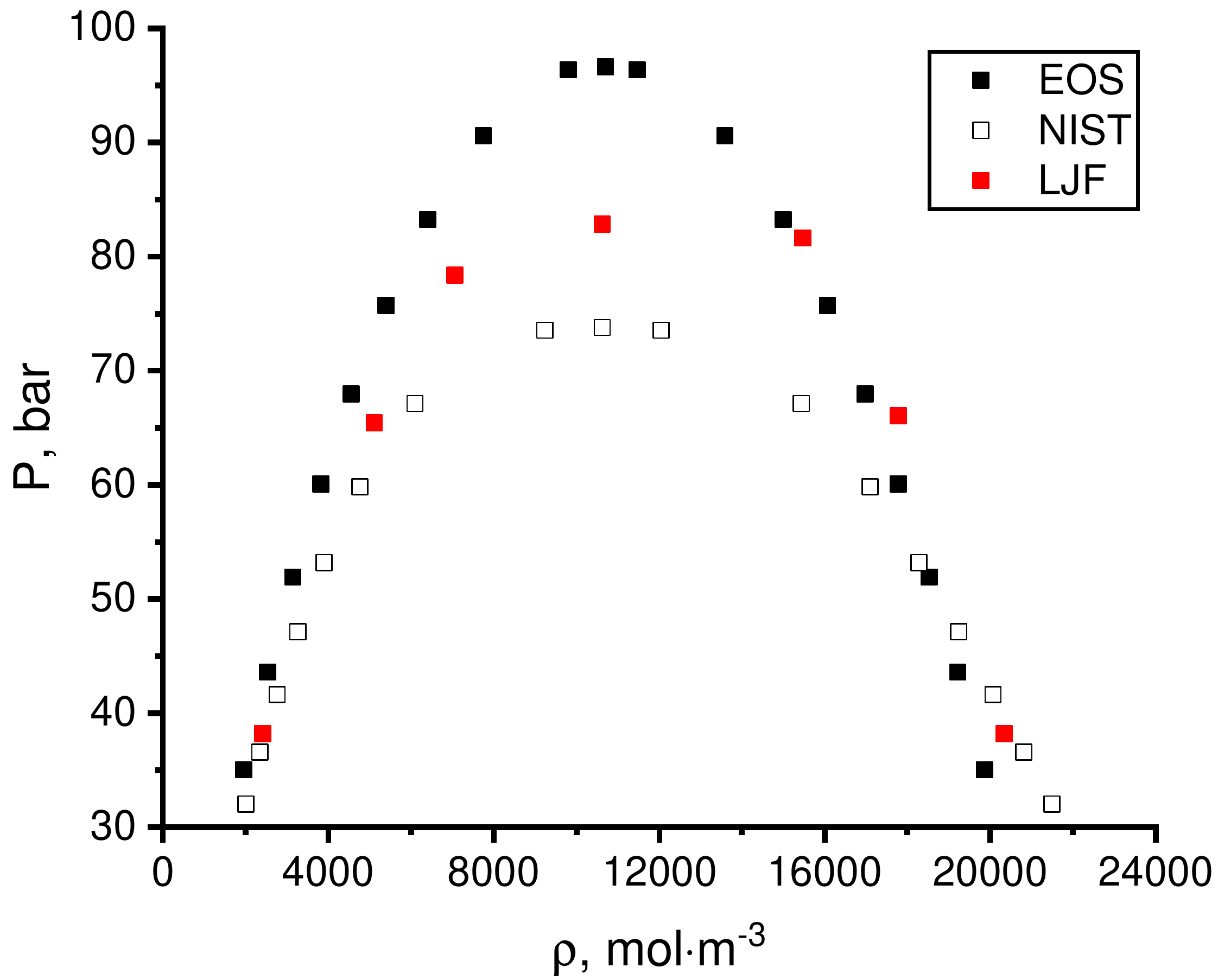}
    \caption{\label{fig01:fig2}}
  \end{subfigure}
  \caption{\textrm{$T-\rho$} (\subref{fig01:fig1}) and \textrm{$P-\rho$} (\subref{fig01:fig2}) coexisting curves of CO$_2$ obtained from the EOS of the cDFT-based approach (black squares), extracted from NIST (empty squares) and taken from the data of the LJ fluid \cite{panagiotopoulos1988phase} (red squares). As it is seen, the model fluids overestimate the critical pressure of the real solvent.}
  \label{fig_01}
\end{figure}

Before turning to the results that we have obtained within the cDFT- and CGMD-based approaches, it is instructive to understand how much our models of solvent deviate from the real one. To understand this, we computed coexistence curves using the equation of state implemented in our cDFT-based approach and compared it to the data of the CO$_2$ phase behavior extracted from NIST and to those obtained by the LJ fluid critical point fitting procedure \cite{potoff1998critical, panagiotopoulos1988phase}. As one can see in the $\mathrm{T}$-$\mathrm{\rho}$ coordinates in Fig.\ref{fig01:fig1} we have a slight deviation of the density at the chosen temperatures, with both models overestimating the vapor branch and underestimating the liquid one. In the case of the $\mathrm{P}$-$\mathrm{\rho}$ coexistence curves in Fig.\ref{fig01:fig2}, it can be seen that the data obtained by the equation of state overestimate the critical pressure by over 20 bar, and the LJ fluid model overestimates it by almost 10 bar. Thus, as we are interested in the supercritical region and the model fluids do not describe the critical pressure accurately, it must be said that the results obtained by the cDFT approach for the pressures lower than approximately 95 bar and by the CGMD technique for the pressures lower than 80 bar may not be reliable. It must be said that the complex molecular potentials utilized in the AAMD simulations, in our case these are the Zhang and TraPPE models, also demonstrate a distinct divergence from the experimental coexistence curves \cite{zhang2005optimized,merker2008comment}. 

\section*{\sffamily \Large Comparison of radial distribution functions}

Another interesting point is the comparison of the structural properties of model fluids. Fig.\ref{fig_02} shows ASP-CO$_2$ radial distribution functions (RDF) computed by the cDFT approach and CGMD simulations with the corresponding potential parameters (see Tables \ref{table01},\ref{table02}). The difference in the location of the first peaks  is approximately 0.2 $\mathrm{\si\angstrom}$, which basically follows from the differences in the parameters of the potential. We also computed the ASP-CO$_2$ center of mass RDF by the AAMD simulations, using the OPLS-AA and GAFF force-fields and Zhang model of CO$_2$. Although the location of most peaks and their heights differ drastically, it is interesting to note that the location of the second peak of the OPLS-AA force field-based RDF is quite similar to that of the first peak of the RDF obtained within the CGMD approach. Considering that the solute molecule during the all-atom calculation is not represented as a sphere, the first peak supposedly corresponds to the case where the CO$_2$ molecules approach the ASP molecule from the sides where the effective distance between the molecules is shorter than the hard sphere effective radius. In such a manner we can conclude that the structural description of our model fluids resembles the all-atom center of mass MD simulations results. As it will be seen from the following results, the accurate description of the thermodynamic properties, namely, the liquid-vapor coexistence curve, is more vital for the correct solvation free energy evaluation than the exact account of the local fluid structure.     

\begin{figure}[h]
\center{\includegraphics[width=0.8\linewidth]{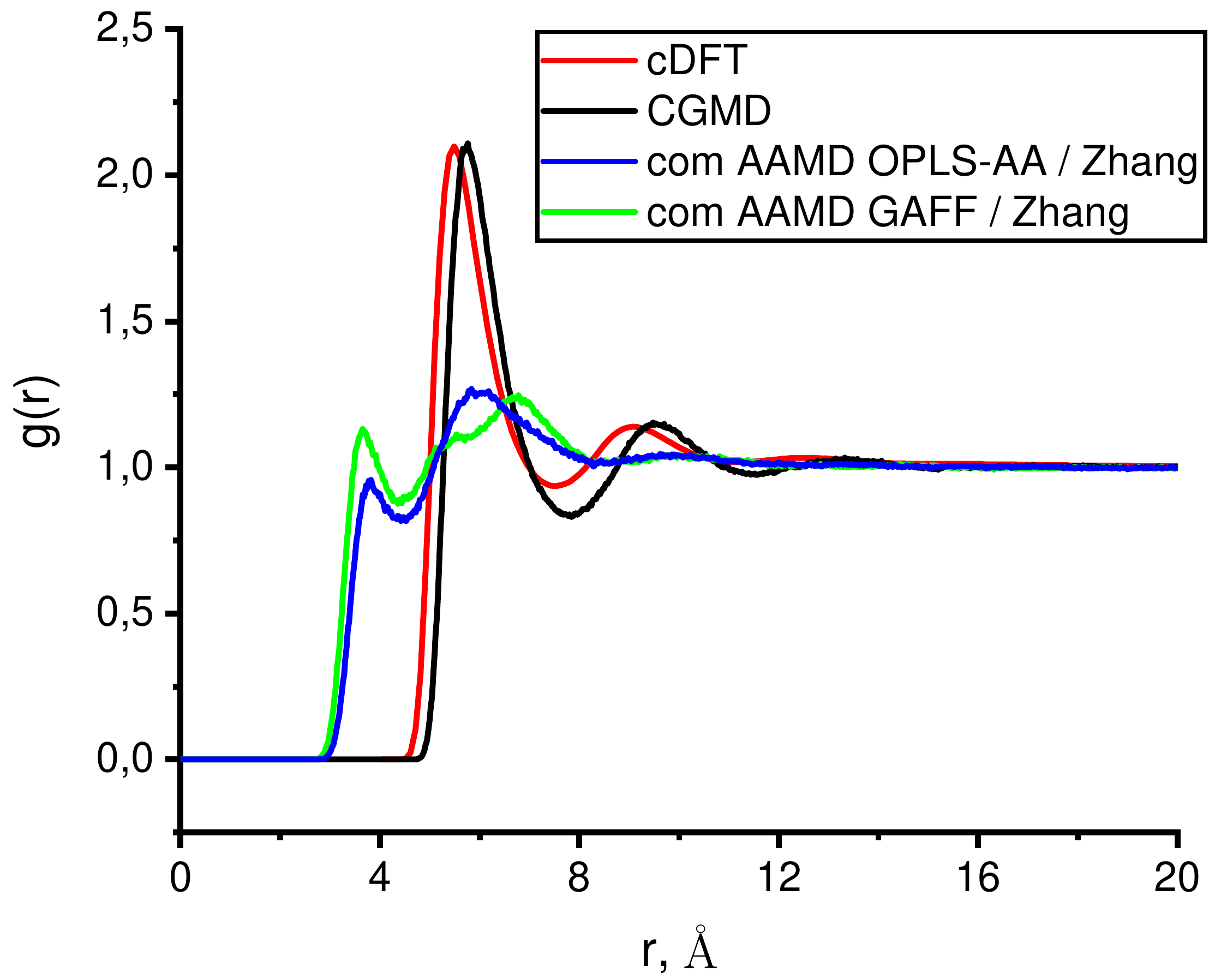}}
\caption{Qualitative comparison of the ASP-CO$_2$ RDFs calculated by the coarse-graining approaches (cDFT - red line, CGMD - black line) with the center-of-mass RDFs computed by the AAMD simulations (force fields: OPLS-AA - blue line and GAFF - green line) for T=308.15 K, P=150 bar.}
\label{fig_02}
\end{figure}

\section*{\sffamily \Large Uncertainties of the AAMD simulations}

\begin{figure}[h]
 \centering
  \begin{subfigure}[b]{0.8\linewidth}
    \centering\includegraphics[width=0.8\linewidth]{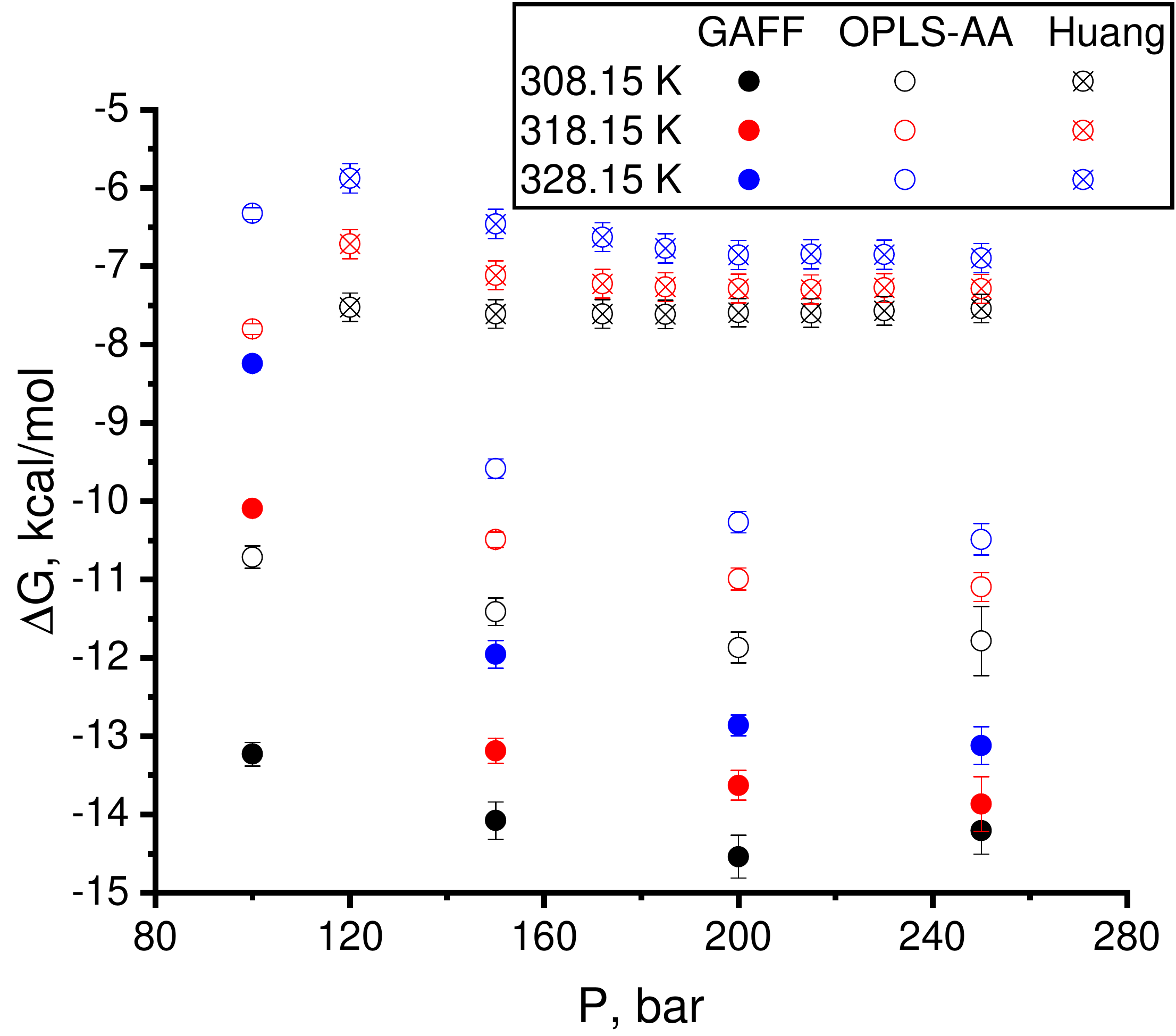}
    \caption{\label{fig1:fig1}}
  \end{subfigure}
  \newline
  \centering
  \begin{subfigure}[b]{0.8\linewidth}
    \centering\includegraphics[width=0.8\linewidth]{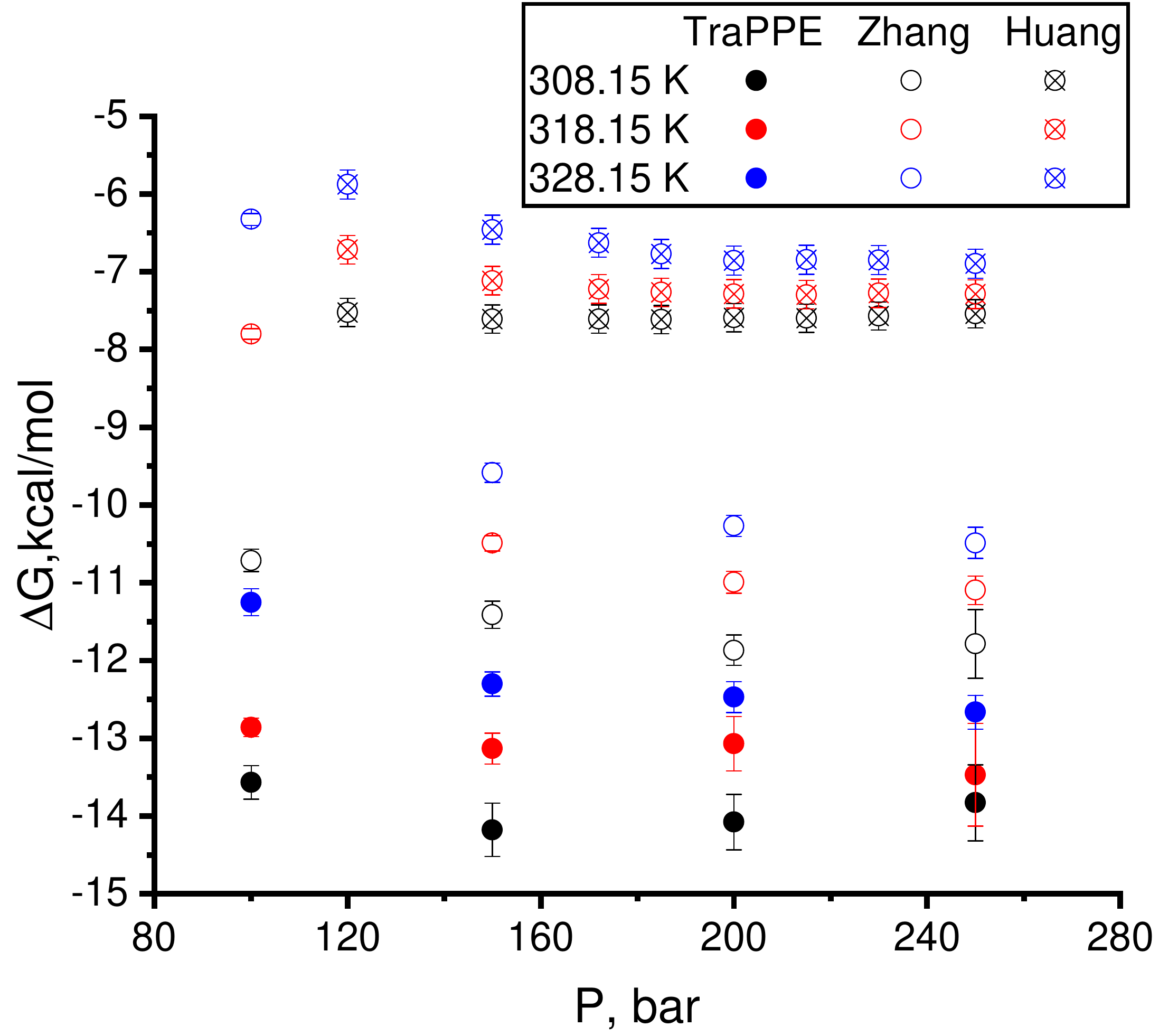}
    \caption{\label{fig1:fig2}}
  \end{subfigure}
  \caption{Comparison of the ASP solvation free energy values for three isotherms obtained from AAMD simulations with OPLS-AA (open circles) and GAFF (filled circles) force fields and the Zhang model for CO$_2$ (\subref{fig1:fig1}) and TraPPE (filled circles) and Zhang (open circles) CO$_2$ models with the OPLS-AA force-field (\subref{fig1:fig2}); experimental data (crossed circles) from \cite{huang2004solubility}.}
  \label{fig_1}
\end{figure}

We would like to underline that the proposed coarse-graining techniques suggest not only the reduction in the time and resources needed to compute the solvation free energy values, but also reduce the ambiguity of the system adjustment prior to the following calculation, as in the case of the AAMD simulation. As we have pointed out above, the results of the solvation free energy calculation using the AAMD simulations heavily rely on the choice of the force field for the solute and solvent molecules \cite{lundborg2015automatic,noroozi2016solvation} and the method of the partial atomic charges calculation \cite{jambeck2013partial}. In this section, we would like to discuss the drawbacks of the AAMD approach we noticed while conducting the simulation for the ASP compound. Fig. \ref{fig1:fig1} shows the differences in the solvation free energy values of the ASP molecule in scCO$_2$ for three isotherms using two force fields: GAFF and OPLS-AA and the Zhang model of CO$_2$ \cite{zhang2005optimized}. As one can see, the discrepancies between the values we have obtained with the different force fields, are quite substantial, growing up to 2 $\mathrm{kcal/mol}$ at most. Fig. \ref{fig1:fig2} shows the aspirin solvation free energy values computed using the OPLS-AA force field and two different solvent models: TraPPE  and Zhang's ones. The qualitative behavior is the same as shown in paper \cite{noroozi2016solvation}, i.e. the values obtained with the TraPPE model are higher than those obtained with the Zhang one, although the discrepancies are much larger. Beside the significant differences in the results obtained using different force fields and models of the system, it is important to note that the absolute values of the solvation free energy we have obtained are quite high compared with the experimental results, indicating rather low possibility of achieving valid results without a thorough preliminary parametrization routine.  

\section*{\sffamily \Large CGMD simulation results}

\begin{figure}[h]
  \centering
  \begin{subfigure}[b]{0.6\linewidth}
    \centering\includegraphics[width=0.9\linewidth]{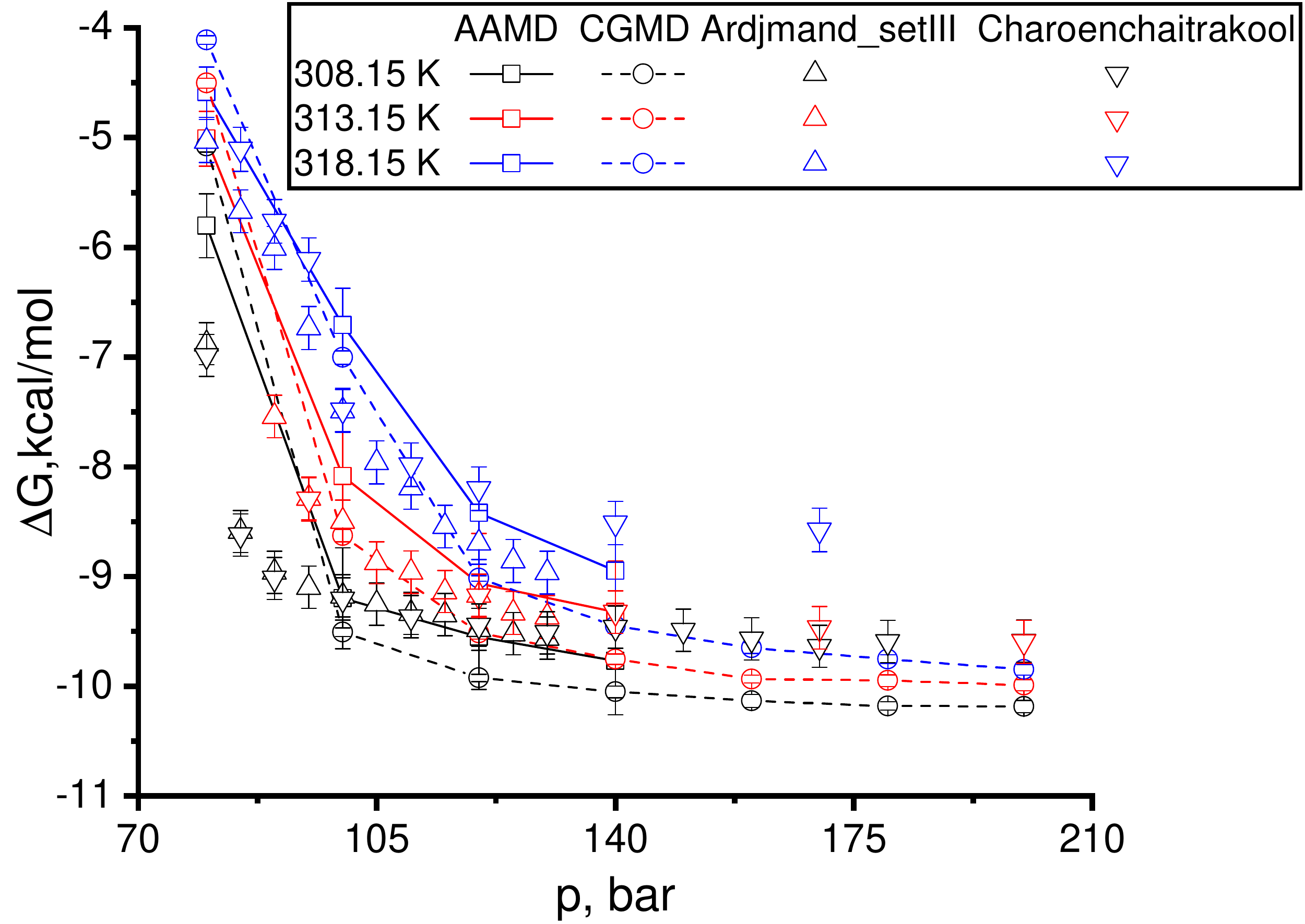}
    \caption{IBU\label{fig02:fig1}}
  \end{subfigure}%
  \hfill
  \centering
  \begin{subfigure}[b]{0.6\linewidth}
    \centering\includegraphics[width=0.9\linewidth]{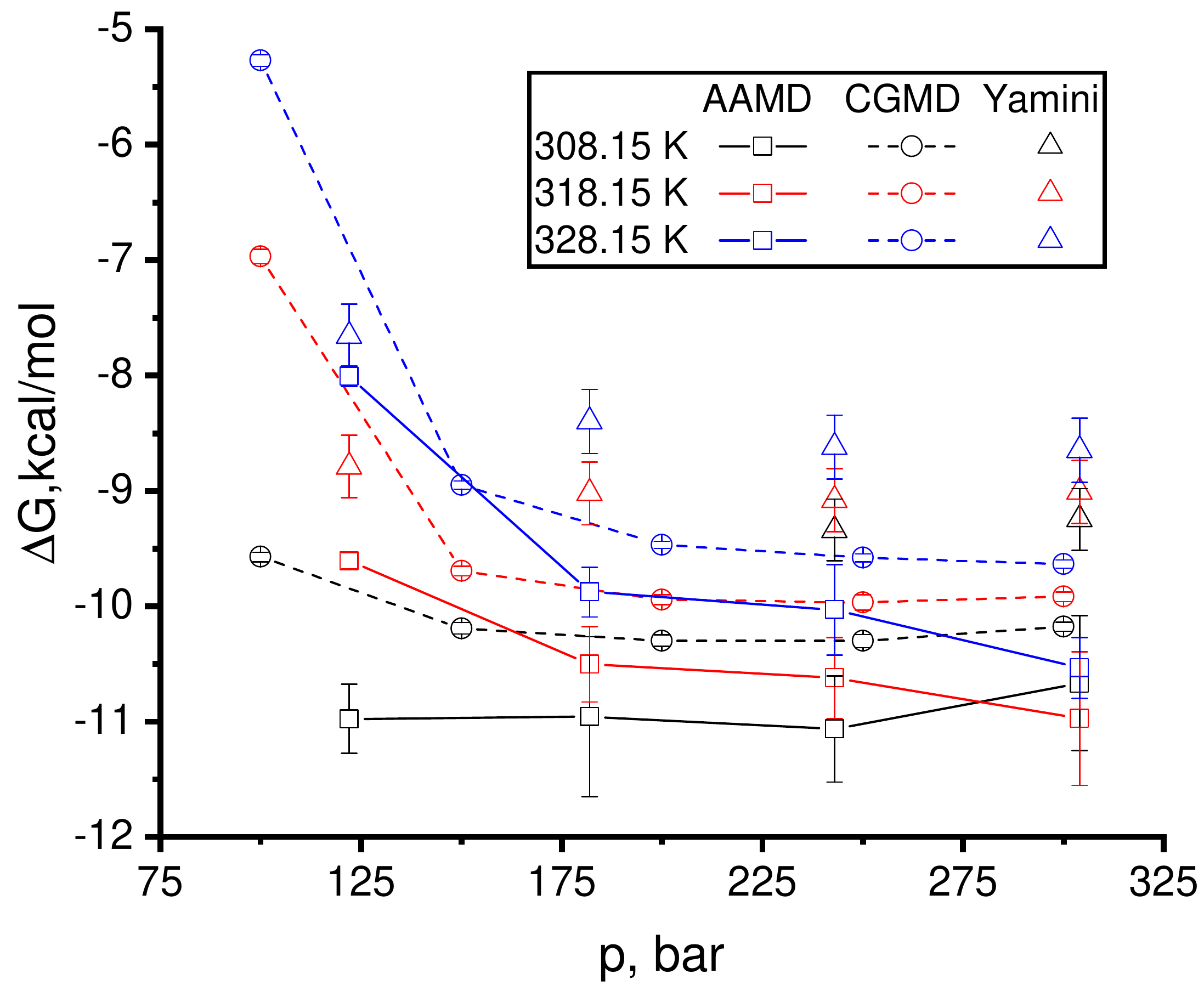}
    \caption{CBZ\label{fig02:fig2}}
  \end{subfigure}
  \newline
   \centering
   \begin{subfigure}[b]{0.6\linewidth}
    \centering\includegraphics[width=0.9\linewidth]{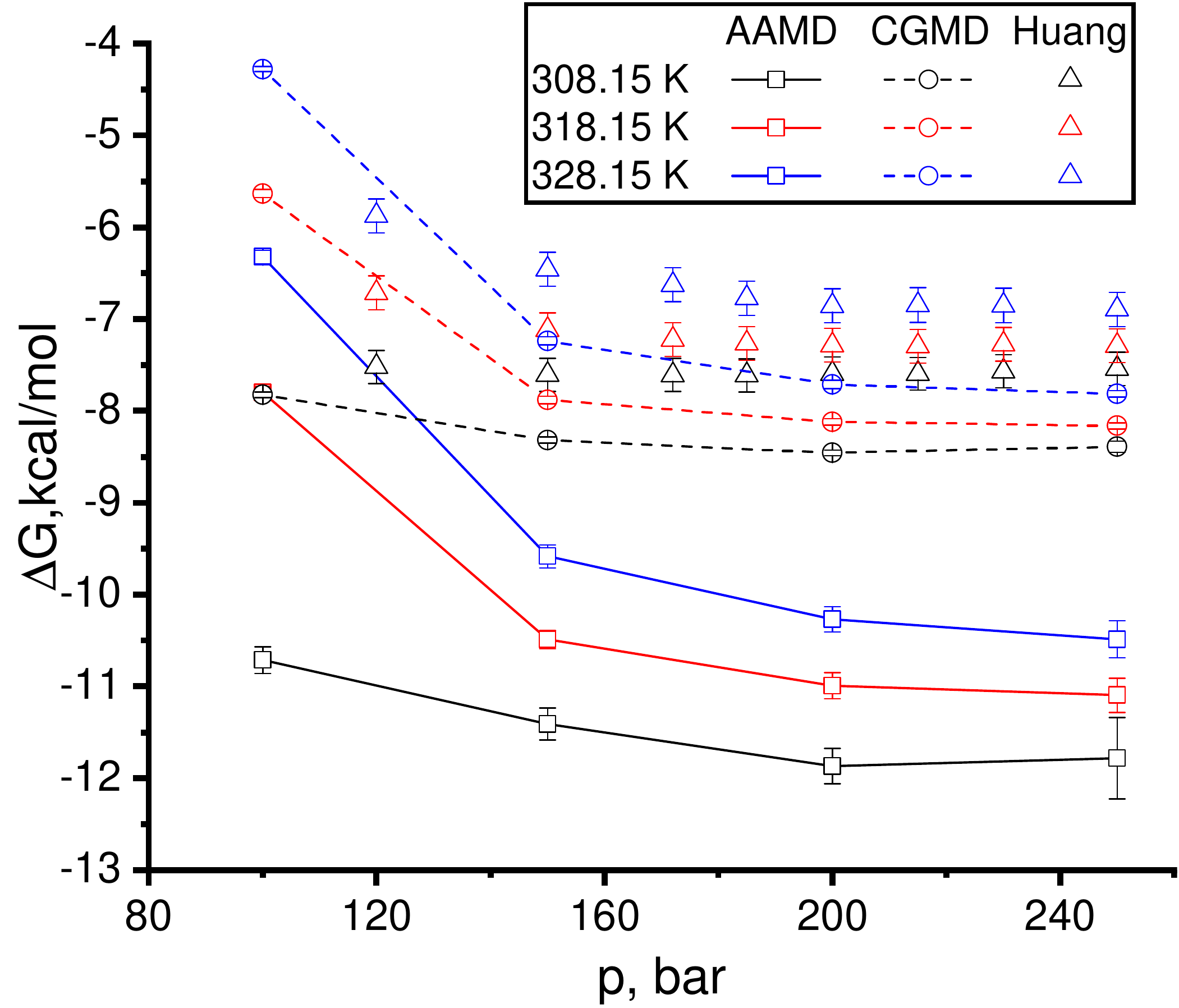}
    \caption{ASP\label{fig02:fig3}}
  \end{subfigure}
  \caption{Comparison of the solvation free energy values obtained within the CGMD approach (open circles), AAMD (open squares) and data extracted from the solubility experiment (open triangles) for IBU, experiment data\cite{charoenchaitrakool2000micronization, ardjmand2014measurement} (\subref{fig02:fig1}); CBZ, experiment data\cite{yamini2001solubilities} (\subref{fig02:fig2}); ASP, experiment data\cite{huang2004solubility} (\subref{fig02:fig3}).}
  \label{fig_02}
\end{figure}

Let us now turn to the results of the proposed CGMD simulation technique. Fig.\ref{fig_02} shows a comparison of the solvation free energy values obtained within the proposed CGMD approach with the interaction parameters presented in Table \ref{table02}, AAMD simulations and the data extracted from the solubility experiments for IBU at 308.15, 313.15 and 318.15 K (\subref{fig02:fig1}), carbamazepine (CBZ) at 308.15, 318.15 and 328.15 K (\subref{fig02:fig2}), and ASP at 308.15, 318.15 and 328.15 K (\subref{fig02:fig3}). The experimental data were taken from the papers (see also table \ref{table10}): IBU \cite{charoenchaitrakool2000micronization, ardjmand2014measurement}, CBZ \cite{yamini2001solubilities} and ASP \cite{huang2004solubility}. Regarding the AAMD simulations for the first solute we used the Zhang model to describe the solvent and the OPLS force field for IBU, which was fitted to quantum chemical simulation in scCO2 \cite{fedorova2016conformational}, thus leading to much better agreement with the experimental data than the two others, where there was no preliminary parameterization; for the second solute we used the generic GROMOS 54A7 force-field \cite{schmid2011definition} with the partial atomic charges of CBZ computed by the Merz-Kollman method \cite{singh1984approach,besler1990atomic}, using the Gaussian 09 software \cite{frisch2013gaussian} with the PBE functional and 6-311++g(2d,p) basis set, which led to better agreement with the experimental data; for the case of ASP we show here the combination of the OPLS solute force field with Zhang's CO$_2$ model as the one providing results that are closest to the experimental data. As one can see, the results are rather surprising in the sense that a crude CGMD-based approach can show reasonably good agreement with the experiment with the largest deviation being near 1 kcal/mol, even outperforming the AAMD procedures.

\section*{\sffamily \Large cDFT results}

\begin{figure}[h]
\center{\includegraphics[width=0.7\linewidth]{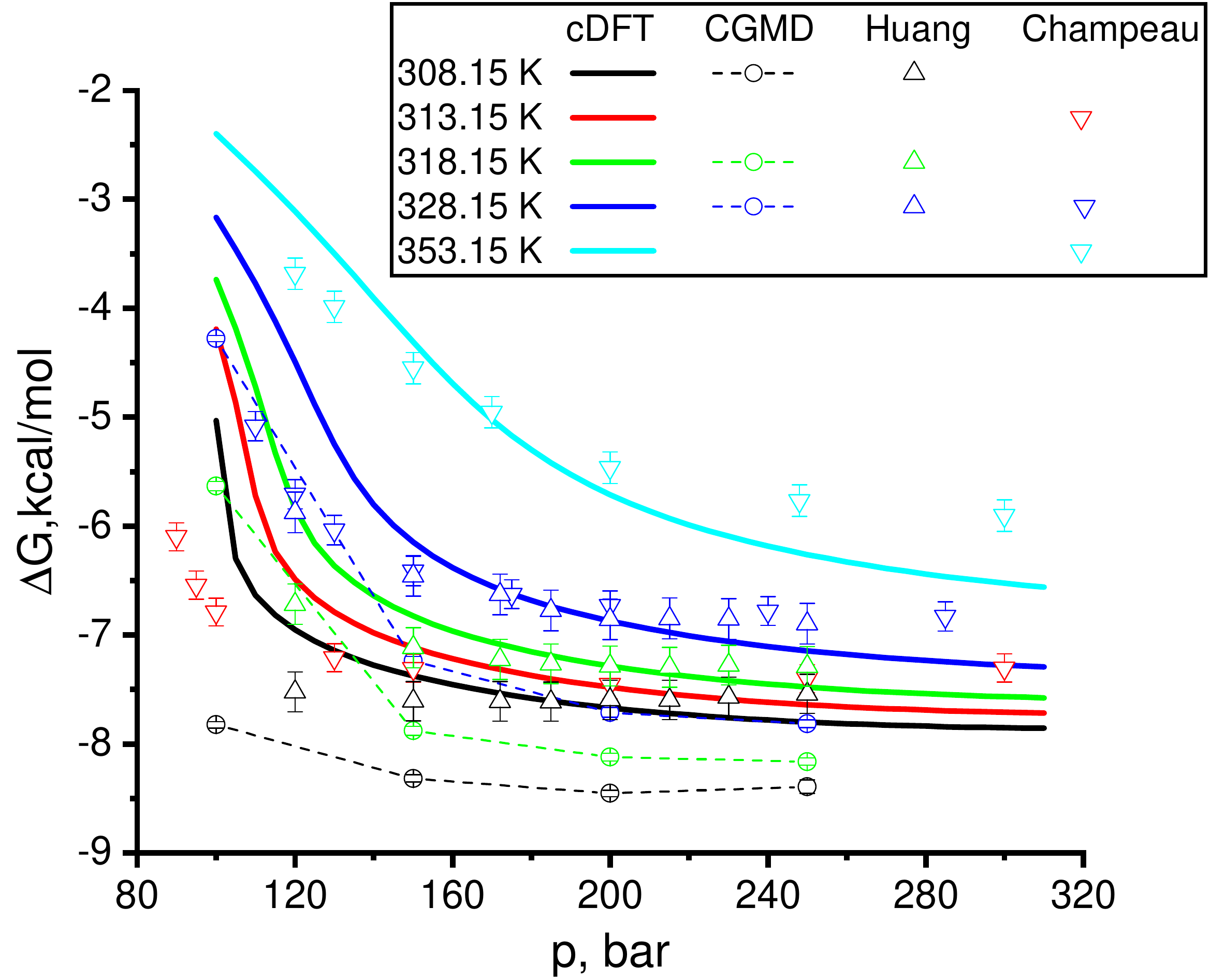}}
\caption{Comparison of the ASP solvation free energy values obtained within the cDFT approach (solid lines), CGMD (open circles) and experimental results (open triangles) \cite{champeau2016solubility,huang2004solubility}.}
  \label{fig_3}
\end{figure}

Now let us compare the results of the described cDFT approach with the results of the experimental measurements and CGMD simulations. Fig.\ref{fig_3} shows the values of the solvation free energy of ASP in scCO$_2$ obtained from the cDFT calculations, CGMD and from the data of the solubility measurement experiments \cite{champeau2016solubility,huang2004solubility}. The ASP molar volume was estimated by the group-contribution method and equals $129.64\times 10^{-6}$ $\mathrm{m^3/mol}$ \cite{cao2008use}. One can observe quite good agreement of the cDFT results with the experimental data, especially in the region of the higher pressure values, while the CGMD results are slightly understated.

\begin{figure}[h]
\center{\includegraphics[width=0.7\linewidth]{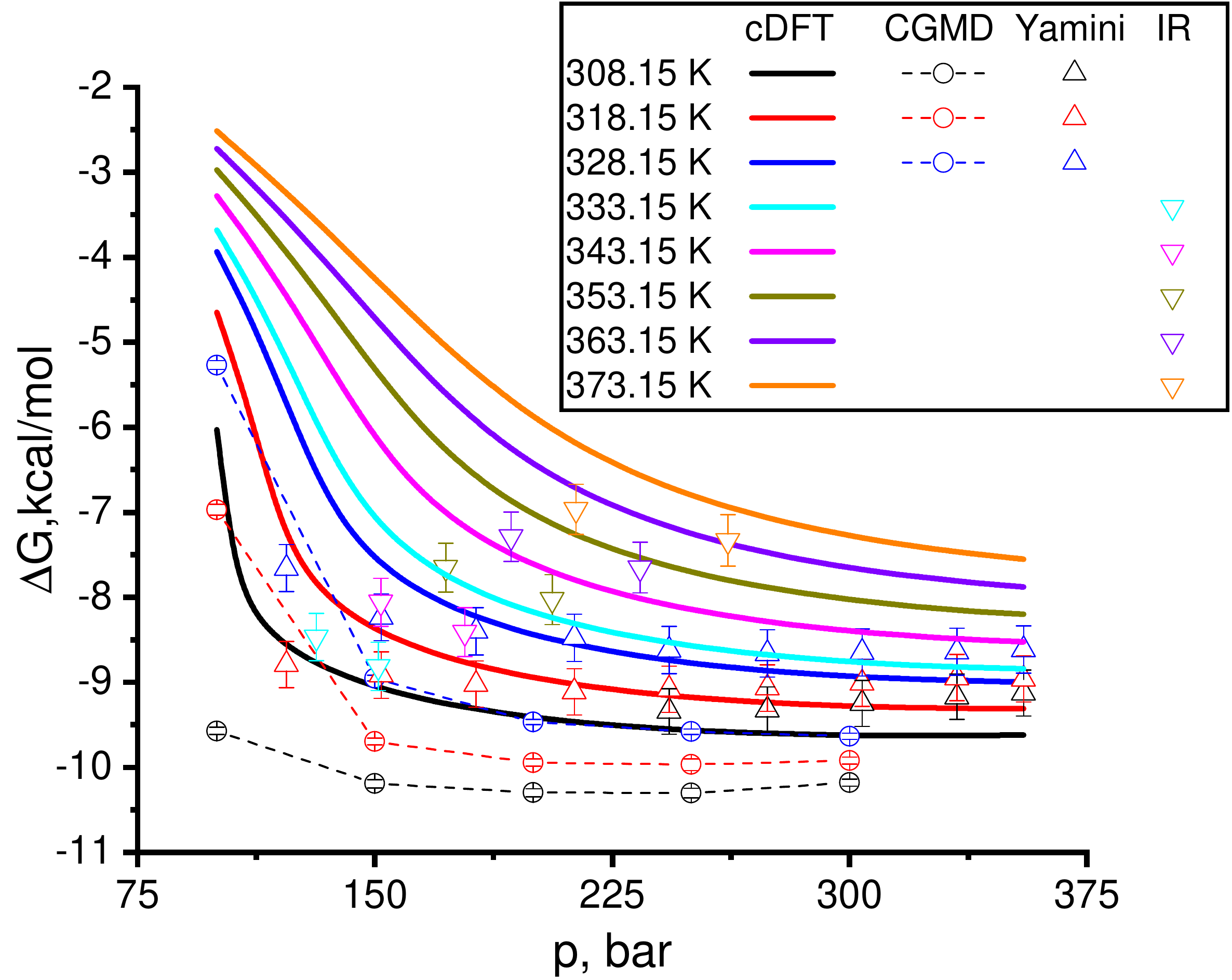}}
 \caption{Comparison of the CBZ solvation free energy values obtained within the cDFT approach (solid lines), CGMD (open circles) and experimental results (open triangles) \cite{yamini2001solubilities, kalikin2020carbamazepine}.}
  \label{fig_4}
\end{figure}

Fig.\ref{fig_4} shows the values of the carbamazepine solvation free energy obtained within the cDFT approach, CGMD and experimental studies \cite{yamini2001solubilities,kalikin2020carbamazepine}. Overall it can be concluded that the agreement is quantitatively decent for the isotherms at 308.15, 318.15 and 328.15 K. The comparison for the isotherms at 333.15-373.15 K does not demonstrate such sufficient agreement with the experiment, although the higher the temperature and the pressure, the better the agreement. Since the infrared (IR) spectroscopy experiment was conducted under the isochoric conditions \cite{kalikin2020carbamazepine}, we cannot propose a straightforward comparison between the two experimental sets. Nevertheless, the quantitative comparison of the solvation free energy values of several closely located points from both sets shows that the IR data exceed Yamini's data set threefold at most.

\begin{figure}[h]
\center{\includegraphics[width=0.7\linewidth]{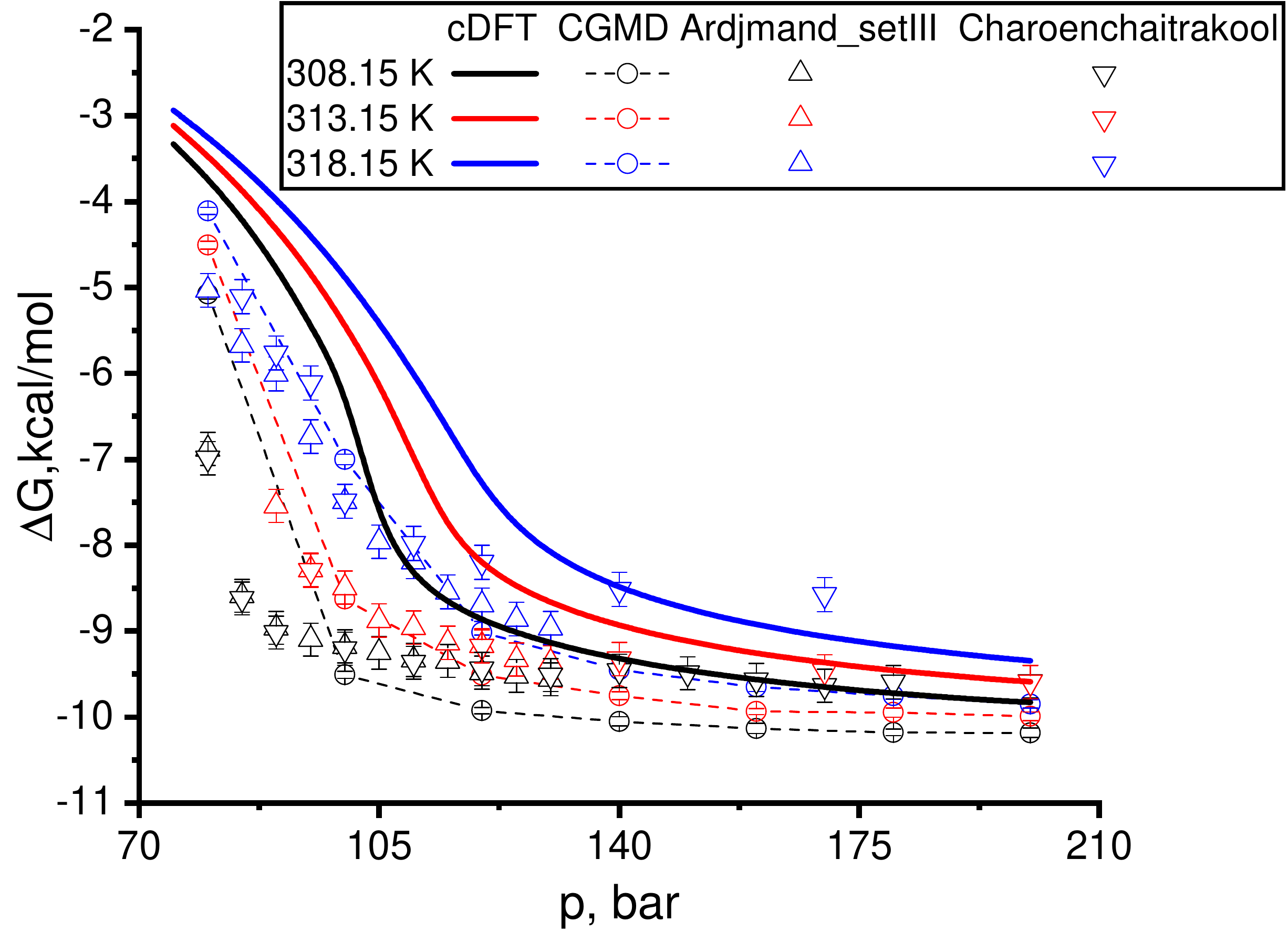}}
\caption{Comparison of the IBU solvation free energy values obtained within the cDFT approach (solid lines), CGMD (open circles) and experimental results (open triangles) \cite{ardjmand2014measurement,charoenchaitrakool2000micronization}.}
\label{fig_5}
\end{figure}

Fig. \ref{fig_5} demonstrates a comparison of the IBU solvation free energy values obtained within the cDFT approach, CGMD simulations and experimental data \cite{ardjmand2014measurement,charoenchaitrakool2000micronization}. Here the critical parameters of IBU were estimated by the Ambrose method. The correspondence with the experimental data is not as pronounced, although it should be noted that the agreement between the cDFT results and the experimental data for all the compounds is quite satisfactory at the pressures from approximately 140 bar. For the case of IBU, the majority of the experimental data points lie in the region below this pressure value. On the other hand, when carefully analyzed, the experimental data for the highest isotherm are rather scattered.

\section*{\sffamily \Large CONCLUSION}
We have proposed two coarse-graining approaches for fast computation of solvation free energies of poorly soluble drug compounds in a scCO$_2$ medium. The first one is based on the utilization of the classical density functional theory and the second one is built upon the use of MD simulations of a supercritical solution represented as a Lennard Jones fluid. The rationalization of such approximation lies in the fact that the main contribution to the interactions between the solute and solvent molecules for the described systems of sparingly soluble compounds in a scCO$_2$ medium is that of the van der Waals type. It seems more crucial to model the fluid phase diagram correctly in this region by the fitting of the critical liquid-vapor point than to make a detailed description of the internal structure of the molecules. The satisfying agreement reached between the obtained results and the experimental ones can be a reason to believe that such approaches can be applied as an acceptable alternative to the all-atom MD simulations.

\appendix

\section{Calculation of the solvation free energy via cDFT approach}

To calculate the solvation Gibbs free energy of the solute in the supercritical fluid, let us start from the expression for the grand thermodynamic potential of scCO$_2$ in the external field with the potential energy $V_{ext}(\mathbf r)$, created by the fixed solute molecule that is posed at the origin
\begin{linenomath}
\begin{equation}\label{3rd_eq}
  \Omega[\rho(\mathbf r)]=F_{int}[\rho(\mathbf r)]+\int d\mathbf r \rho(\mathbf r)(V_{ext}(\mathbf r)-\mu),
\end{equation}
\end{linenomath}
where $F_{int}[\rho(\mathbf r)]$ is the intrinsic Helmholtz free energy of the scCO$_2$ and $\mu$ is the chemical potential of bulk fluid at certain temperature and pressure.
The intrinsic Helmholtz free energy, in its turn, can be comprised of two contributions
\begin{linenomath}
\begin{equation}\label{4th_eq}
  F_{int}[\rho(\mathbf r)]=k_BT\int d\mathbf r\rho(\mathbf r)[\ln(\Lambda^3\rho(\mathbf r))-1]+F_{ex}[\rho(\mathbf r)],
\end{equation}
\end{linenomath}
where the first one is the Helmholtz free energy of the ideal gas and the second one is the excess Helmholtz free energy of fluid; $\Lambda$ is the thermal de Broglie wavelength. We assume the CO$_2$ molecules as coarse-grained hard spheres interacting thorough the effective LJ pairwise potential
\begin{linenomath}
\begin{equation}
V_{ff}(r)=4\varepsilon_{ff}\left(\left(\frac{\sigma_{ff}}{r}\right)^{12}-\left(\frac{\sigma_{ff}}{r}\right)^6\right)\Theta(r_{c}-r),
\end{equation}
\end{linenomath}
where the $\varepsilon_{ff}$ and $\sigma_{ff}$ are LJ parameters of the interaction between two molecules of the fluid and $\Theta(x)$ is the Heaviside step function. The potential is divided into two parts at the minimum of the LJ potential $r_m=2^{1/6}\sigma_{ff}$ in accordance with the WCA procedure \cite{andersen1971relationship}, and cut at $r_c=5\sigma_{ff}$. Thus, the excess part of the free energy can be written as a sum of two contributions as follows
\begin{linenomath}
\begin{equation}\label{5th_eq}
  F_{ex}[\rho(\mathbf r)]=F_{hs}[\rho(\mathbf r)]+F_{att}[\rho(\mathbf r)],
\end{equation}
\end{linenomath}
where the $F_{hs}[\rho(\mathbf r)]$ is the Helmholtz free energy of the hard sphere system and $F_{att}[\rho(\mathbf r)]$ is the contribution of the attractive interaction between the molecules of the fluid.
In its turn, the hard sphere part of the excess free energy is approximated with the help of Rosenfeld's version of the Fundamental Measure Theory (FMT) \cite{rosenfeld1989free} in a certain way
\begin{linenomath}
\begin{equation}\label{6th_eq}
F_{hs}[\rho(\mathbf r)]=k_BT\int d\mathbf r\Phi(\mathbf r),
\end{equation}
\end{linenomath}
where the $\Phi(\mathbf r)$ is the excess free energy density, which is given by the following expression
\begin{linenomath}
\begin{equation}\label{7th_eq}
 \Phi=-n_0\ln(1-n_3)+\frac{n_1n_2- {n_{V1}}\cdot{n_{V2}}}{1-n_3}+\frac{n_2^3-3n_2{n_{V1}}\cdot{n_{V2}}}{24\pi(1-n_3)^2},
\end{equation}
\end{linenomath}
where $n_i\cdot n_j$ denotes scalar multiplication.

The excess Helmholtz free energy density $\Phi(\mathbf r)$ is a function of the weighted densities
\begin{linenomath}
\begin{equation}\label{8th_eq}
n_{\alpha}(\mathbf{r})=\int d\mathbf{r}'\rho(\mathbf{r}')\omega^{(\alpha)}(\mathbf{r}-\mathbf{r}'),
\end{equation}
\end{linenomath}
where the $\omega^{(\alpha)}(\mathbf{r}-\mathbf{r}')$ are the weighted functions, which characterize the hard sphere geometry: the volume, the surface area and the mean radius of the curvature. There are six such functions for the case of the three-dimensional hard sphere system, but only three of them are independent \cite{rosenfeld1989free}
\begin{linenomath}
\begin{align}
\omega^{(3)}(\mathbf{r})&=\Theta(R-r)\nonumber\\
\omega^{(2)}(\mathbf{r})&=\delta(R-r)\\
{\omega}^{(V2)}(\mathbf{r})&=\frac{\mathbf{r}}{r}\delta(R-r)\nonumber.
\end{align}
\end{linenomath}
Other three are dependent on the described above functions in the following way: $\omega^{(1)}=\omega^{(2)}/4\pi R$, $\omega^{(0)}=\omega^{(2)}/4\pi R^2$, $\omega^{(V1)}=\omega^{(V2)}/4\pi R$, 
where $\delta(r)$ is the Dirac delta-function and $R=d_{BH}/2$ is the effective hard sphere radius. The effective Barker-Henderson (BH) diameter is determined by the following Pade approximation \cite{Verlet1972a}
\begin{linenomath}
\begin{equation}
d_{BH} = \sigma_{ff}\frac{1.068 \varepsilon_{ff}/k_B T + 0.3837}{\varepsilon_{ff}/k_B T + 0.4293}.
\end{equation}
\end{linenomath}
The weighted functions give rise to the corresponding weighted densities $n_{\alpha}(\mathbf{r})$
\begin{linenomath}
\begin{align}
  n_{3}(r)&=\frac{\pi}{r}\int\limits_{r-R}^{r+R} dr' r' \rho(r')[R^2 - (r - r')^2]\nonumber\\
  n_{2}(r)&=\frac{2\pi R}{r}\int\limits_{r-R}^{r+R} dr' r' \rho(r')\\
  n_{V2}(r)&=\frac{\pi}{r^2}\int\limits_{r-R}^{r+R} dr' r' \rho(r') [r^2 + R^2 - r'^2]\nonumber
\end{align}
\end{linenomath}
The contribution of the interparticle attraction to the excess free energy is determined by the mean-field approximation
\begin{linenomath}
\begin{equation}\label{11th_eq}
  F_{att}[\rho(\mathbf{r})]=\frac{1}{2}\int d\mathbf{r}\rho(\mathbf{r})\int d\mathbf{r}'\rho(\mathbf{r}')\phi_{WCA}(\mathbf{r}-\mathbf{r}'),
\end{equation}
\end{linenomath}
where the effective WCA pair potential of attractive interactions is
\begin{linenomath}
\begin{equation}\label{12th_eq}
  \phi_{WCA}(r) = \left\{
  \begin{array}{lr}
    -\varepsilon_{ff}, & r<r_m \\
    4\varepsilon_{ff}\left[\left(\frac{\sigma_{ff}}{r}\right)^{12}-\left(\frac{\sigma_{ff}}{r}\right)^{6}\right], & r_m<r<r_c.
  \end{array}
  \right.
\end{equation}
\end{linenomath}
We model the solute molecule as a center generating external LJ potential
\begin{linenomath}
\begin{equation}
V_{ext}(\mathbf{r})=4 \varepsilon_{sf}\left(\left(\frac{\sigma_{sf}}{r}\right)^{12}-\left(\frac{\sigma_{sf}}{r}\right)^6\right)
\end{equation}
\end{linenomath}
with the effective parameters of interaction between the solute and fluid molecules. We obtain these parameters with the help of the Berthelot-Lorenz mixing rules: $\sigma_{sf}=(\sigma_{ss}+\sigma_{ff})/2$ and $\varepsilon_{sf}=\sqrt{\varepsilon_{ss}\varepsilon_{ff}}$, where the parameters of the interaction between two molecules of the active compound ($\sigma_{ss}$, $\varepsilon_{ss}$) and CO$_2$ ($\sigma_{ff}$, $\varepsilon_{ff}$) can be obtained from the fitting of the respective critical parameters\footnote{Critical density and temperature - for CO$_2$, critical pressure and temperature - for the compound.} of the liquid-gas transition, which values are available in the literature (see below). 

For the bulk phase of the fluid, i.e. when $V_{ext}(\mathbf{r})=0$ and $\rho(\mathbf{r})=\rho_b=const$, the functional (\ref{4th_eq}) can be reduced to the following form
\begin{linenomath}
\begin{equation}
f=\frac{F_{int}}{V}=f_{id}+f_{ex},
\end{equation}
\end{linenomath}
where $V$ is the system volume, $f_{id}=\rho_b k_BT(\ln(\Lambda^3\rho_b)-1)$ is the ideal free energy density, and the excess contribution to the Helmholtz free energy density of the bulk fluid takes the following form
\begin{linenomath}
\begin{equation}
    f_{ex}(\rho_b,T)=\rho_b k_BT\left(-\ln(1-\eta)+\frac{3\eta}{1-\eta}+\frac{3\eta^2}{2(1-\eta)^2}\right)+\frac{1}{2}B_{WCA}\rho_b^2,
\end{equation}
\end{linenomath}
where $\eta=\pi d_{BH}^3\rho_b/6$ is the hard sphere system packing fraction and the following auxiliary function, corresponding to the attractive contribution, is introduced
\begin{linenomath}
\begin{equation}
    B_{WCA}=-\frac{32\sqrt2}{9}\pi\varepsilon_{ff}\sigma_{ff}^3+\frac{16}{3}\pi\varepsilon_{ff}\sigma_{ff}^3\left[\left(\frac{\sigma_{ff}}{r_c}\right)^3-\frac{1}{3}\left(\frac{\sigma_{ff}}{r_c}\right)^9\right].
\end{equation}
\end{linenomath}
The total pressure then takes the following form
\begin{linenomath}
\begin{equation}
    P=\rho_b\frac{\partial f}{\partial\rho_b}-f=\rho_{b} k_BT\frac{1+\eta+\eta^2}{(1-\eta)^3}+\frac{1}{2}B_{WCA}\rho_b^2.
\end{equation}
\end{linenomath}
One can obtain LJ parameters of interaction between the fluid molecules as a result of solving following system of the equations, using known critical parameters of CO$_2$
\begin{linenomath}
\begin{equation}
    \frac{\partial P}{\partial\rho_b}\Bigr|_{\substack{T=T_c\\\rho_b=\rho_c}}=0, ~~ \frac{\partial^2 P}{\partial\rho_b^2}\Bigr|_{\substack{T=T_c\\\rho_b=\rho_c}}=0.
\end{equation}
\end{linenomath}
In the same way the LJ parameters for the solute can be found.

In order to obtain the density profile, it is necessary to minimize the grand thermodynamic potential with respect to $\rho(\mathbf{r})$ and solve numerically the Euler-Lagrange equation
\begin{linenomath}
\begin{equation}\label{9th_eq}
  \frac{\delta\Omega[\rho(\mathbf{r})]}{\delta\rho(\mathbf{r})}=0,
\end{equation}
\end{linenomath}
this leads to the following expression
\begin{linenomath}
\begin{equation}\label{10th_eq}
  \rho(\mathbf{r})=\rho_b\exp\left[\frac{\mu_{ex}(\rho_b,T)-c^{(1)}_{fmt}(\mathbf{r})+\int d\mathbf{r}'\rho(\mathbf{r}')\phi_{WCA}(\mathbf{r}-\mathbf{r}')-V_{ext}(\mathbf{r})}{k_BT}\right],
\end{equation}
\end{linenomath}
where $\mu_{ex}(\rho_b,T)$ is the excess chemical potential of the bulk phase, $c^{(1)}_{fmt}(\mathbf{r})=\delta F_{hs}[\rho(\mathbf{r})]/{\delta\rho(\mathbf{r})}$ is the one-particle direct correlation function of the hard-sphere system within the FMT. From the definition of the one-particle direct correlation function and from the obtained expression for the excess free energy within the FMT, we can write the following
\begin{linenomath}
\begin{equation}
c^{(1)}_{fmt}(\mathbf{r}) = k_{B}T \int\limits_{r - R}^{r+R} dr^{\prime}\left[\frac{\pi}{r}\frac{\partial \Phi}{\partial n_3}r^{\prime}(R^2 - (r - r')^2) + \frac{2\pi R}{r}\frac{\partial \Phi}{\partial n_2} r^{\prime}+ \frac{\partial \Phi}{\partial n_{V2}}\frac{\pi}{r} (r'^2 - r^2 + R^2) \right], \end{equation}
\end{linenomath}
where the derivatives are determined as follows
\begin{linenomath}
\begin{align}
\frac{\partial \Phi}{\partial n_3} &= \frac{n_2}{4\pi R^2(1 - n_3)} + \frac{n_2^2 - n_{V2}^2}{4\pi R (1 - n_3)^2} + \frac{n_2^3 - 3n_2 n_{V2}^2}{12\pi(1-n_3)^3},\nonumber\\
\frac{\partial \Phi}{\partial n_2} &= -\ln(1 - n_3)/(4\pi R^2) + \frac{n_2}{2\pi R(1 - n_3)} + \frac{n_2^2 - n_{V2}^2}{8\pi(1-n_3)^2},\\
\frac{\partial \Phi}{\partial n_{V2}} &= -\frac{n_{V2}}{2\pi R(1 - n_3)} - \frac{n_2 n_{V2}}{4\pi(1-n_3)^2}\nonumber.    
\end{align}
\end{linenomath}

In order to calculate the contribution of the attractive interactions between two fluid particles
in spherical geometry, one have to distinguish the situations where $r - r_c < 0$ and $r - r_c > 0$. Also, using the spherical symmetry of the system, we can calculate the integral only for $\mathbf{r} = r \mathbf{z}$ without loss of generality. Here we will present the equations for the latter case as an example
\begin{linenomath}
\begin{eqnarray}
\int d\mathbf{r}'\rho(\mathbf{r}')\phi_{WCA}(\mathbf{r}-\mathbf{r}') &=&  2\pi \varepsilon_{ff} \int_{r - r_c}^{r-r_m}d{r}' {r}' \rho({r}') \frac{1}{r} \biggl[ \frac{2\sigma_{ff}^{12}}{5(r - {r}')^{10}} - \frac{\sigma_{ff}^6}{(r - {r}')^4} - \frac{2\sigma_{ff}^{12}}{5r_c^{10}} + \frac{\sigma_{ff}^6}{r_c^4} \biggr]\nonumber \\
&+& 2\pi \varepsilon_{ff} \int_{r + r_m}^{r+r_c}d{r}' {r}' \rho({r}') \frac{1}{r} \biggl[ \frac{2\sigma_{ff}^{12}}{5(r - {r}')^{10}} - \frac{\sigma_{ff}^6}{(r - {r}')^4} - \frac{2\sigma_{ff}^{12}}{5r_c^{10}} + \frac{\sigma_{ff}^6}{r_c^4} \biggr] \nonumber \\
&-&\frac{\pi \varepsilon_{ff}}{r} \int_{r-r_m}^{r+r_m} d{r}' {r}' \rho({r}') [r_m^2 - (r - {r}')^2] \nonumber \\
&+& 2\pi \varepsilon_{ff} \int_{r - r_m}^{r+r_m}d{r}' {r}' \rho({r}') \frac{1}{r} \biggl[ \frac{2\sigma_{ff}^{12}}{5r_m^{10}} - \frac{\sigma_{ff}^6}{r_m^4} - \frac{2\sigma_{ff}^{12}}{5r_c^{10}} + \frac{\sigma_{ff}^6}{r_c^4} \biggr].
\end{eqnarray}
\end{linenomath}

Finally, the solvation Gibbs free energy can be calculated as the excess grand thermodynamic potential as follows
\begin{linenomath}
\begin{equation}
  \Delta G_{solv} = \Omega[\rho(\mathbf r)] - \Omega[\rho_b].
\end{equation}
\end{linenomath}

\section{Aspirin partial atomic charges}
\begin{figure}[h]
  
  \begin{subfigure}[b]{1\linewidth}
    \centering
    \centering\includegraphics[width=1\linewidth]{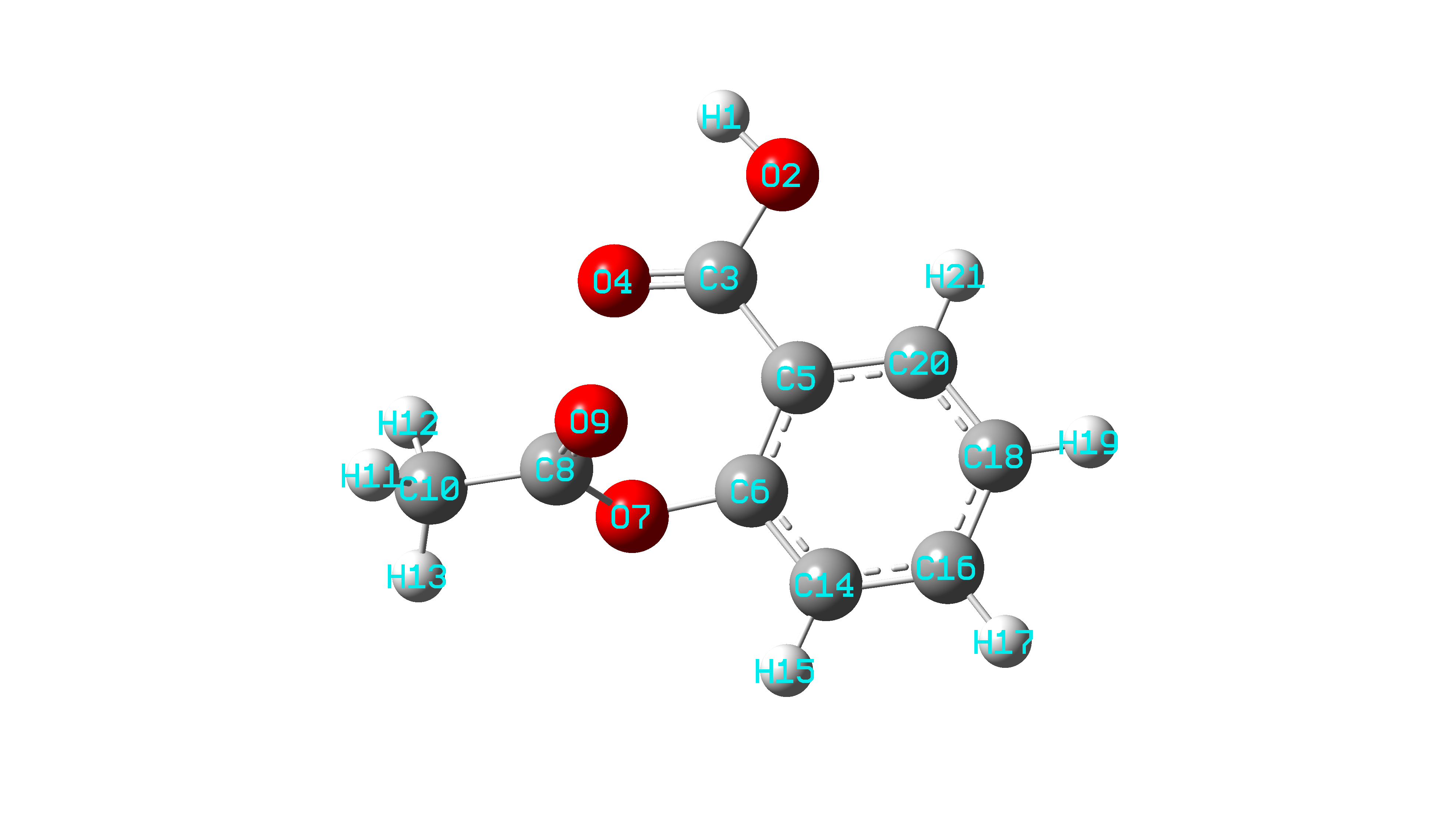}
    \caption{\label{fig_app1:fig1}}
  \end{subfigure}%
  \newline
  \begin{subfigure}[b]{1\linewidth}
    \centering
    \centering\includegraphics[width=1\linewidth]{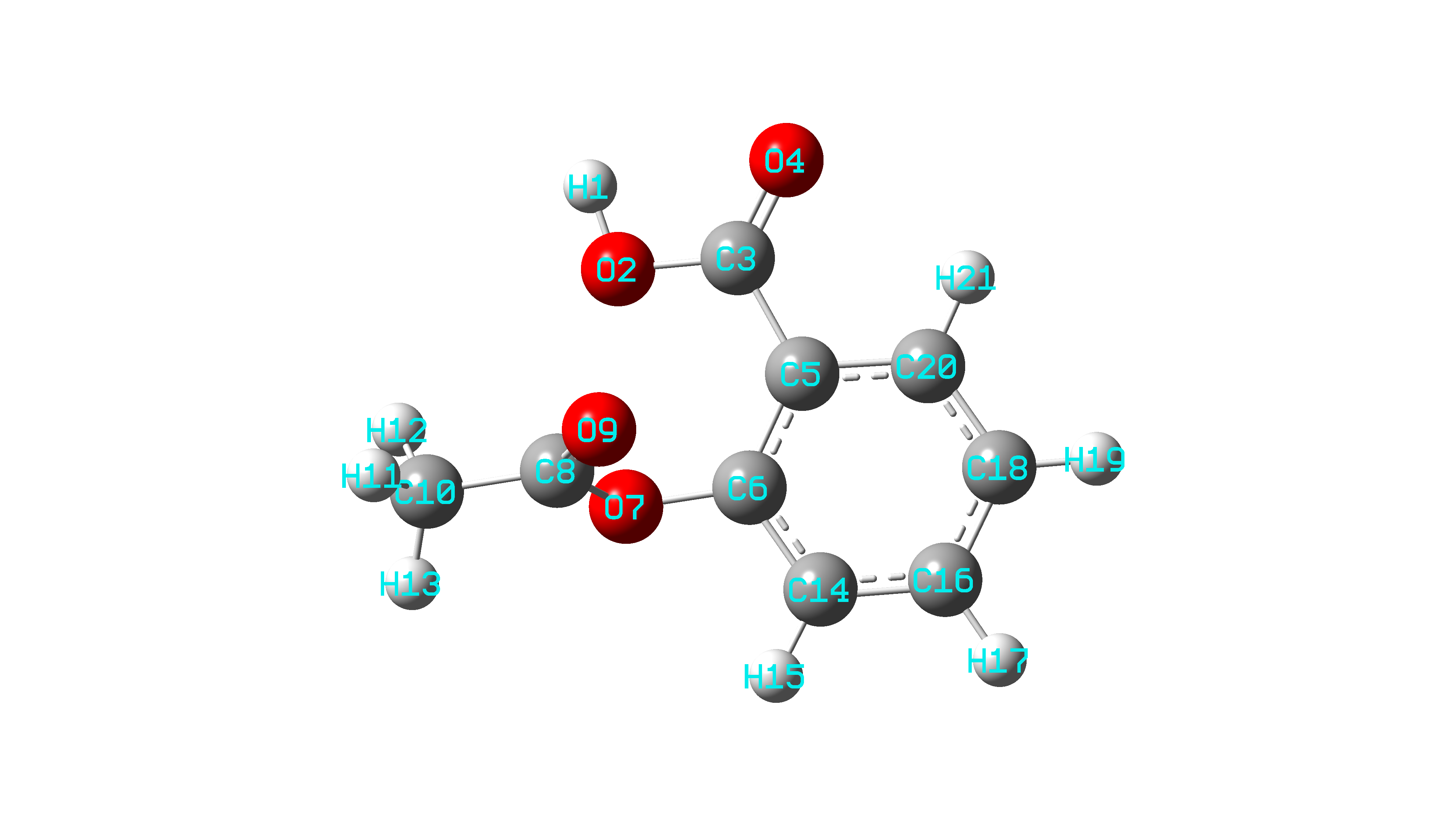}
    \caption{\label{fig_app1:fig2}}
  \end{subfigure}
  \caption{Two most stable ASP conformers as calculated from quantum chemistry.}
  \label{fig_app1}
\end{figure}

Two most stable ASP conformers are represented at the Fig. \ref{fig_app1}. The partial atomic charges, averaged over these two conformers, are shown in Table \ref{table20}.

\begin{table}[h!]
\centering
\caption{Atomic partial charges of the ASP conformers, computed with the Merz-Kollman method, PBE functional, 6-311++g(2d,p) basis set.}
\begin{tabular}{l|r|l|r}
\multicolumn{2}{c}{conf.1} & \multicolumn{2}{|c}{conf.2} \\
      \hline
       atom & charge & atom & charge \\
      \hline
     H1  &     0.402490  &  H1  &     0.394904  \\
     O2  &    -0.551911  &  O2  &    -0.530321  \\
     C3  &     0.632537  &  C3  &     0.613256  \\
     O4  &    -0.513812  &  O4  &    -0.503629  \\
     C5  &    -0.106338  &  C5  &    -0.118470  \\
     C6  &     0.374670  &  C6  &     0.396578  \\
     O7  &    -0.416109  &  O7  &    -0.418618  \\
     C8  &     0.768870  &  C8  &     0.772999  \\
     O9  &    -0.502975  &  O9  &    -0.500559  \\
     C10 &    -0.557154  &  C10 &    -0.600964  \\
     H11 &     0.152293  &  H11 &     0.166754  \\
     H12 &     0.194728  &  H12 &     0.192129  \\
     H13 &     0.155861  &  H13 &     0.172574  \\
     C14 &    -0.259618  &  C14 &    -0.278193  \\
     H15 &     0.165669  &  H15 &     0.169621  \\
     C16 &    -0.081902  &  C16 &    -0.083413  \\
     H17 &     0.136997  &  H17 &     0.139358  \\
     C18 &    -0.156598  &  C18 &    -0.161393  \\
     H19 &     0.136687  &  H19 &     0.139205  \\
     C20 &    -0.137291  &  C20 &    -0.116117  \\
     H21 &     0.162907  &  H21 &     0.154300  \\
     \end{tabular}
 \label{table20}
\end{table}

\section{Experimental solvation free energy data}

\includepdf[pages=-]{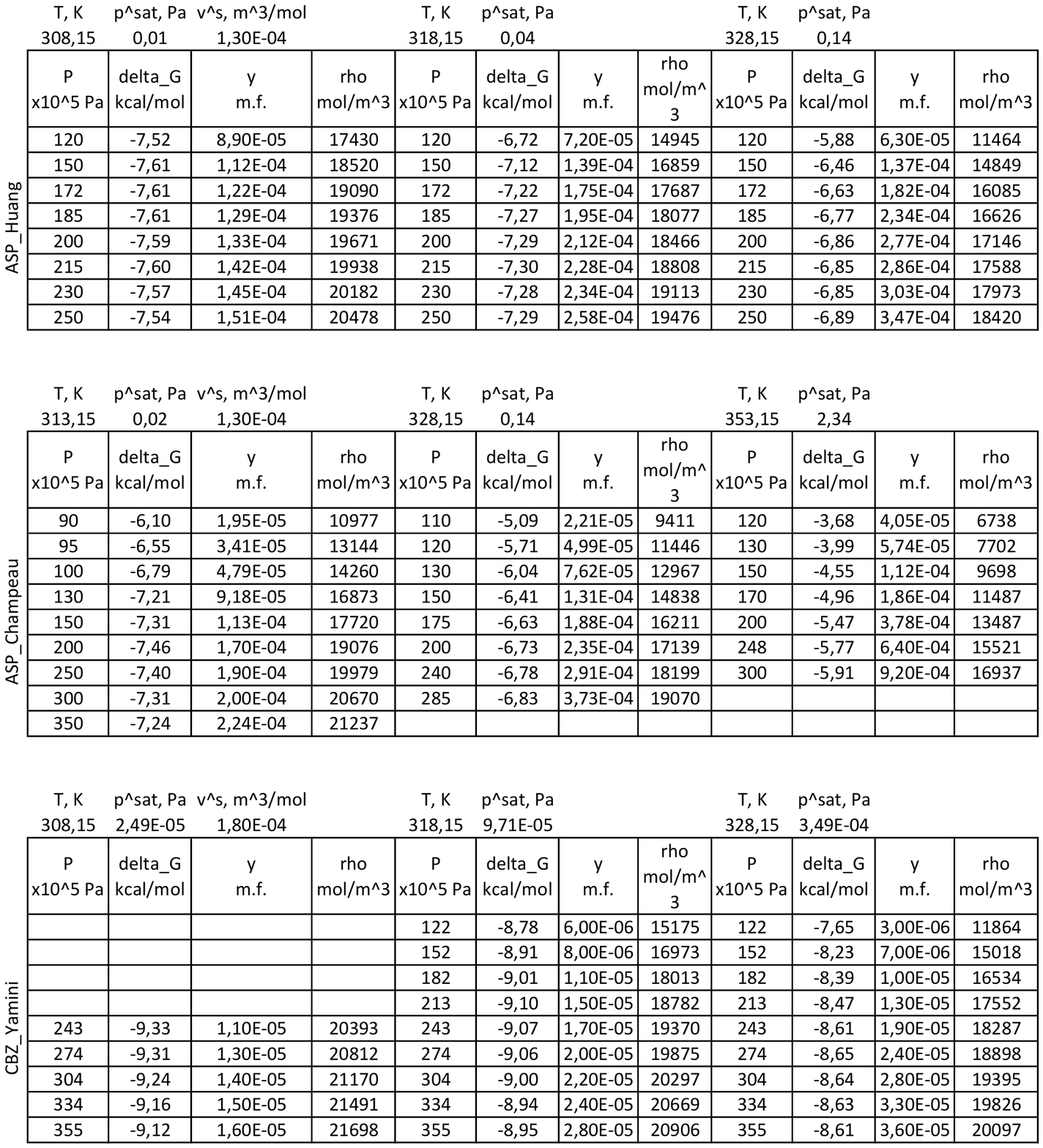}

\subsection*{\sffamily \large ACKNOWLEDGMENTS}

The research was funded by the Ministry of Science and Higher Education of the Russian Federation (grant no. RFMEFI61618$\times$0097). This research was done on the supercomputer facilities provided by NRU HSE.

\bibliography{literature}

\begin{thebibliography}{10}

\bibitem{noroozi2016solvation}
J.~Noroozi, C.~Ghotbi, J.~J. Sardroodi, J.~Karimi-Sabet, and M.~A. Robert,
  ``Solvation free energy and solubility of acetaminophen and ibuprofen in
  supercritical carbon dioxide: Impact of the solvent model,'' {\em The Journal
  of Supercritical Fluids}, vol.~109, pp.~166--176, 2016.

\bibitem{frolov2015accurate}
A.~I. Frolov, ``Accurate calculation of solvation free energies in
  supercritical fluids by fully atomistic simulations: Probing the theory of
  solutions in energy representation,'' {\em Journal of chemical theory and
  computation}, vol.~11, no.~5, pp.~2245--2256, 2015.

\bibitem{bruckner2011efficiency}
S.~Bruckner and S.~Boresch, ``Efficiency of alchemical free energy simulations.
  ii. improvements for thermodynamic integration,'' {\em Journal of
  computational chemistry}, vol.~32, no.~7, pp.~1320--1333, 2011.

\bibitem{hansen2014practical}
N.~Hansen and W.~F. Van~Gunsteren, ``Practical aspects of free-energy
  calculations: a review,'' {\em Journal of chemical theory and computation},
  vol.~10, no.~7, pp.~2632--2647, 2014.

\bibitem{jia2016calculations}
X.~Jia, M.~Wang, Y.~Shao, G.~K\"{o}nig, B.~R. Brooks, J.~Z. Zhang, and Y.~Mei,
  ``Calculations of solvation free energy through energy reweighting from
  molecular mechanics to quantum mechanics,'' {\em Journal of chemical theory
  and computation}, vol.~12, no.~2, pp.~499--511, 2016.

\bibitem{misin2016hydration}
M.~Misin, M.~V. Fedorov, and D.~S. Palmer, ``Hydration free energies of
  molecular ions from theory and simulation,'' {\em The Journal of Physical
  Chemistry B}, vol.~120, no.~5, pp.~975--983, 2016.

\bibitem{shirts2012best}
M.~R. Shirts, ``Best practices in free energy calculations for drug design,''
  in {\em Computational drug discovery and design}, pp.~425--467, Springer,
  2012.

\bibitem{lundborg2015automatic}
M.~Lundborg and E.~Lindahl, ``Automatic gromacs topology generation and
  comparisons of force fields for solvation free energy calculations,'' {\em
  The Journal of Physical Chemistry B}, vol.~119, no.~3, pp.~810--823, 2015.

\bibitem{jambeck2013partial}
J.~P. J{\"a}mbeck, F.~Mocci, A.~P. Lyubartsev, and A.~Laaksonen, ``Partial
  atomic charges and their impact on the free energy of solvation,'' {\em
  Journal of computational chemistry}, vol.~34, no.~3, pp.~187--197, 2013.

\bibitem{da2020all}
G.~C. da~Silva, G.~M. Silva, F.~W. Tavares, F.~P. Fleming, and B.~A. Horta,
  ``Are all-atom any better than united-atom force fields for the description
  of liquid properties of alkanes?,'' {\em Journal of Molecular Modeling},
  vol.~26, no.~11, pp.~1--17, 2020.

\bibitem{glova2019toward}
A.~D. Glova, I.~V. Volgin, V.~M. Nazarychev, S.~V. Larin, S.~V. Lyulin, and
  A.~A. Gurtovenko, ``Toward realistic computer modeling of paraffin-based
  composite materials: critical assessment of atomic-scale models of
  paraffins,'' {\em RSC Advances}, vol.~9, no.~66, pp.~38834--38847, 2019.

\bibitem{papavasileiou2019molecular}
K.~D. Papavasileiou, L.~D. Peristeras, A.~Bick, and I.~G. Economou, ``Molecular
  dynamics simulation of pure n-alkanes and their mixtures at elevated
  temperatures using atomistic and coarse-grained force fields,'' {\em The
  Journal of Physical Chemistry B}, vol.~123, no.~29, pp.~6229--6243, 2019.

\bibitem{ewen2016comparison}
J.~P. Ewen, C.~Gattinoni, F.~M. Thakkar, N.~Morgan, H.~A. Spikes, and D.~Dini,
  ``A comparison of classical force-fields for molecular dynamics simulations
  of lubricants,'' {\em Materials}, vol.~9, no.~8, p.~651, 2016.

\bibitem{garlapati2009temperature}
C.~Garlapati and G.~Madras, ``Temperature independent mixing rules to correlate
  the solubilities of antibiotics and anti-inflammatory drugs in scco2,'' {\em
  Thermochimica Acta}, vol.~496, no.~1-2, pp.~54--58, 2009.

\bibitem{moine2019can}
E.~Moine, R.~Privat, J.-N. Jaubert, B.~Sirjean, N.~Novak, E.~Voutsas, and
  C.~Boukouvalas, ``Can we safely predict solvation gibbs energies of pure and
  mixed solutes with a cubic equation of state?,'' {\em Pure and Applied
  Chemistry}, vol.~91, no.~8, pp.~1295--1307, 2019.

\bibitem{kontogeorgis2020equations}
G.~M. Kontogeorgis, X.~Liang, A.~Arya, and I.~Tsivintzelis, ``Equations of
  state in three centuries. are we closer to arriving to a single model for all
  applications?,'' {\em Chemical Engineering Science: X}, vol.~7, p.~100060,
  2020.

\bibitem{hutacharoen2017predicting}
P.~Hutacharoen, S.~Dufal, V.~Papaioannou, R.~M. Shanker, C.~S. Adjiman,
  G.~Jackson, and A.~Galindo, ``Predicting the solvation of organic compounds
  in aqueous environments: from alkanes and alcohols to pharmaceuticals,'' {\em
  Industrial \& Engineering Chemistry Research}, vol.~56, no.~38,
  pp.~10856--10876, 2017.

\bibitem{el2013application}
H.~A.~A. El, C.~Si-Moussa, S.~Hanini, and M.~Laidi, ``Application of pc-saft
  and cubic equations of state for the correlation of solubility of some
  pharmaceutical and statin drugs in sc-co2,'' {\em Chemical Industry and
  Chemical Engineering Quarterly/CICEQ}, vol.~19, no.~3, pp.~449--460, 2013.

\bibitem{anvari2014study}
M.~H. Anvari and G.~Pazuki, ``A study on the predictive capability of the
  saft-vr equation of state for solubility of solids in supercritical co2,''
  {\em The Journal of Supercritical Fluids}, vol.~90, pp.~73--83, 2014.

\bibitem{yang2005modeling}
H.~Yang and C.~Zhong, ``Modeling of the solubility of aromatic compounds in
  supercritical carbon dioxide--cosolvent systems using saft equation of
  state,'' {\em The Journal of supercritical fluids}, vol.~33, no.~2,
  pp.~99--106, 2005.

\bibitem{sodeifian2020experimental}
G.~Sodeifian, S.~A. Sajadian, and R.~Derakhsheshpour, ``Experimental
  measurement and thermodynamic modeling of lansoprazole solubility in
  supercritical carbon dioxide: Application of saft-vr eos,'' {\em Fluid Phase
  Equilibria}, vol.~507, p.~112422, 2020.

\bibitem{mahmoudabadi2021application}
S.~Z. Mahmoudabadi and G.~Pazuki, ``Application of pc-saft eos for
  pharmaceuticals: Solubility, co-crystal, and thermodynamic modeling,'' {\em
  Journal of Pharmaceutical Sciences}, 2021.

\bibitem{ramirez2020parameterization}
N.~Ram{\'\i}rez-V{\'e}lez, A.~Pi\~na Martinez, J.-N. Jaubert, and R.~Privat,
  ``Parameterization of saft models: Analysis of different parameter estimation
  strategies and application to the development of a comprehensive database of
  pc-saft molecular parameters,'' {\em Journal of Chemical \& Engineering
  Data}, vol.~65, no.~12, pp.~5920--5932, 2020.

\bibitem{shimoyama2009development}
Y.~Shimoyama and Y.~Iwai, ``Development of activity coefficient model based on
  cosmo method for prediction of solubilities of solid solutes in supercritical
  carbon dioxide,'' {\em The Journal of Supercritical Fluids}, vol.~50, no.~3,
  pp.~210--217, 2009.

\bibitem{wang2014predictive}
L.-H. Wang and S.-T. Lin, ``A predictive method for the solubility of drug in
  supercritical carbon dioxide,'' {\em The Journal of Supercritical Fluids},
  vol.~85, pp.~81--88, 2014.

\bibitem{marenich2009universal}
A.~V. Marenich, C.~J. Cramer, and D.~G. Truhlar, ``Universal solvation model
  based on solute electron density and on a continuum model of the solvent
  defined by the bulk dielectric constant and atomic surface tensions,'' {\em
  The Journal of Physical Chemistry B}, vol.~113, no.~18, pp.~6378--6396, 2009.

\bibitem{chamberlin2008modeling}
A.~C. Chamberlin, D.~G. Levitt, C.~J. Cramer, and D.~G. Truhlar, ``Modeling
  free energies of solvation in olive oil,'' {\em Molecular pharmaceutics},
  vol.~5, no.~6, pp.~1064--1079, 2008.

\bibitem{klamt2005cosmo}
A.~Klamt, {\em COSMO-RS: from quantum chemistry to fluid phase thermodynamics
  and drug design}.
\newblock Elsevier, 2005.

\bibitem{misin2016predicting}
M.~Misin, D.~S. Palmer, and M.~V. Fedorov, ``Predicting solvation free energies
  using parameter-free solvent models,'' {\em The Journal of Physical Chemistry
  B}, vol.~120, no.~25, pp.~5724--5731, 2016.

\bibitem{sokolov2007fundamental}
V.~F. Sokolov and G.~N. Chuev, ``Fundamental measure theory of hydrated
  hydrocarbons,'' {\em Journal of molecular modeling}, vol.~13, no.~2,
  pp.~319--326, 2007.

\bibitem{chuev2006hydration}
G.~Chuev and V.~Sokolov, ``Hydration of hydrophobic solutes treated by the
  fundamental measure approach,'' {\em The Journal of Physical Chemistry B},
  vol.~110, no.~37, pp.~18496--18503, 2006.

\bibitem{rosenfeld1989free}
Y.~Rosenfeld, ``Free-energy model for the inhomogeneous hard-sphere fluid
  mixture and density-functional theory of freezing,'' {\em Physical review
  letters}, vol.~63, no.~9, p.~980, 1989.

\bibitem{peng1976new}
D.-Y. Peng and D.~B. Robinson, ``A new two-constant equation of state,'' {\em
  Industrial \& Engineering Chemistry Fundamentals}, vol.~15, no.~1,
  pp.~59--64, 1976.

\bibitem{zhao2011molecular}
S.~Zhao, R.~Ramirez, R.~Vuilleumier, and D.~Borgis, ``Molecular density
  functional theory of solvation: From polar solvents to water,'' {\em The
  Journal of chemical physics}, vol.~134, no.~19, p.~194102, 2011.

\bibitem{zhao2011new}
S.~Zhao, Z.~Jin, and J.~Wu, ``New theoretical method for rapid prediction of
  solvation free energy in water,'' {\em The Journal of Physical Chemistry B},
  vol.~115, no.~21, pp.~6971--6975, 2011.

\bibitem{sergiievskyi2014fast}
V.~P. Sergiievskyi, G.~Jeanmairet, M.~Levesque, and D.~Borgis, ``Fast
  computation of solvation free energies with molecular density functional
  theory: Thermodynamic-ensemble partial molar volume corrections,'' {\em The
  journal of physical chemistry letters}, vol.~5, no.~11, pp.~1935--1942, 2014.

\bibitem{gendre2009classical}
L.~Gendre, R.~Ramirez, and D.~Borgis, ``Classical density functional theory of
  solvation in molecular solvents: Angular grid implementation,'' {\em Chemical
  Physics Letters}, vol.~474, no.~4-6, pp.~366--370, 2009.

\bibitem{baghbanbashi2020solubility}
M.~Baghbanbashi, N.~Hadidi, and G.~Pazuki, ``Solubility of pharmaceutical
  compounds in supercritical carbon dioxide: Application, experimental, and
  mathematical modeling,'' in {\em Green Sustainable Process for Chemical and
  Environmental Engineering and Science}, pp.~185--254, Elsevier, 2020.

\bibitem{padrela2018supercritical}
L.~Padrela, M.~A. Rodrigues, A.~Duarte, A.~M. Dias, M.~E. Braga, and H.~C.
  de~Sousa, ``Supercritical carbon dioxide-based technologies for the
  production of drug nanoparticles/nanocrystals--a comprehensive review,'' {\em
  Advanced drug delivery reviews}, vol.~131, pp.~22--78, 2018.

\bibitem{archer2017standard}
A.~J. Archer, B.~Chacko, and R.~Evans, ``The standard mean-field treatment of
  inter-particle attraction in classical dft is better than one might expect,''
  {\em The Journal of chemical physics}, vol.~147, no.~3, p.~034501, 2017.

\bibitem{Verlet1972a}
L.~Verlet and J.-J. Weis, ``Equilibrium theory of simple liquids,'' {\em
  Physical Review A}, vol.~5, no.~2, p.~939, 1972.

\bibitem{nist}
E.~W. Lemmon, M.~O. McLinden, and D.~G. Friend, {\em Thermophysical properties
  of fluid systems}.
\newblock NIST chemistry WebBook, 1998.

\bibitem{malde2011automated}
A.~K. Malde, L.~Zuo, M.~Breeze, M.~Stroet, D.~Poger, P.~C. Nair,
  C.~Oostenbrink, and A.~E. Mark, ``An automated force field topology builder
  (atb) and repository: version 1.0,'' {\em Journal of chemical theory and
  computation}, vol.~7, no.~12, pp.~4026--4037, 2011.

\bibitem{stroet2018automated}
M.~Stroet, B.~Caron, K.~M. Visscher, D.~P. Geerke, A.~K. Malde, and A.~E. Mark,
  ``Automated topology builder version 3.0: prediction of solvation free
  enthalpies in water and hexane,'' {\em Journal of chemical theory and
  computation}, vol.~14, no.~11, pp.~5834--5845, 2018.

\bibitem{wang2004development}
J.~Wang, R.~M. Wolf, J.~W. Caldwell, P.~A. Kollman, and D.~A. Case,
  ``Development and testing of a general amber force field,'' {\em Journal of
  computational chemistry}, vol.~25, no.~9, pp.~1157--1174, 2004.

\bibitem{kaminski2001evaluation}
G.~A. Kaminski, R.~A. Friesner, J.~Tirado-Rives, and W.~L. Jorgensen,
  ``Evaluation and reparametrization of the opls-aa force field for proteins
  via comparison with accurate quantum chemical calculations on peptides,''
  {\em The Journal of Physical Chemistry B}, vol.~105, no.~28, pp.~6474--6487,
  2001.

\bibitem{zhang2005optimized}
Z.~Zhang and Z.~Duan, ``An optimized molecular potential for carbon dioxide,''
  {\em The Journal of chemical physics}, vol.~122, no.~21, p.~214507, 2005.

\bibitem{potoff2001vapor}
J.~J. Potoff and J.~I. Siepmann, ``Vapor--liquid equilibria of mixtures
  containing alkanes, carbon dioxide, and nitrogen,'' {\em AIChE journal},
  vol.~47, no.~7, pp.~1676--1682, 2001.

\bibitem{canzar2013charge}
S.~Canzar, M.~El-Kebir, R.~Pool, K.~Elbassioni, A.~K. Malde, A.~E. Mark, D.~P.
  Geerke, L.~Stougie, and G.~W. Klau, ``Charge group partitioning in
  biomolecular simulation,'' {\em Journal of Computational Biology}, vol.~20,
  no.~3, pp.~188--198, 2013.

\bibitem{koziara2014testing}
K.~B. Koziara, M.~Stroet, A.~K. Malde, and A.~E. Mark, ``Testing and validation
  of the automated topology builder (atb) version 2.0: prediction of hydration
  free enthalpies,'' {\em Journal of computer-aided molecular design}, vol.~28,
  no.~3, pp.~221--233, 2014.

\bibitem{budkov2019possibility}
Y.~Budkov, A.~Kolesnikov, D.~Ivlev, N.~Kalikin, and M.~Kiselev, ``Possibility
  of pressure crossover prediction by classical dft for sparingly dissolved
  compounds in scco2,'' {\em Journal of Molecular Liquids}, vol.~276,
  pp.~801--805, 2019.

\bibitem{kalikin2020carbamazepine}
N.~Kalikin, M.~Kurskaya, D.~Ivlev, M.~Krestyaninov, R.~Oparin, A.~Kolesnikov,
  Y.~Budkov, A.~Idrissi, and M.~Kiselev, ``Carbamazepine solubility in
  supercritical co2: A comprehensive study,'' {\em Journal of Molecular
  Liquids}, p.~113104, 2020.

\bibitem{pronk2013gromacs}
S.~Pronk, S.~P{\'a}ll, R.~Schulz, P.~Larsson, P.~Bjelkmar, R.~Apostolov, M.~R.
  Shirts, J.~C. Smith, P.~M. Kasson, D.~van~der Spoel, {\em et~al.}, ``Gromacs
  4.5: a high-throughput and highly parallel open source molecular simulation
  toolkit,'' {\em Bioinformatics}, vol.~29, no.~7, pp.~845--854, 2013.

\bibitem{abraham2015gromacs}
M.~J. Abraham, T.~Murtola, R.~Schulz, S.~P{\'a}ll, J.~C. Smith, B.~Hess, and
  E.~Lindahl, ``Gromacs: High performance molecular simulations through
  multi-level parallelism from laptops to supercomputers,'' {\em SoftwareX},
  vol.~1, pp.~19--25, 2015.

\bibitem{bekker1993gromacs}
H.~Bekker, H.~Berendsen, E.~Dijkstra, S.~Achterop, R.~Vondrumen,
  D.~Vanderspoel, A.~Sijbers, H.~Keegstra, and M.~Renardus, ``Gromacs-a
  parallel computer for molecular-dynamics simulations,'' in {\em 4th
  International Conference on Computational Physics (PC 92)}, pp.~252--256,
  World Scientific Publishing, 1993.

\bibitem{berendsen1995gromacs}
H.~J. Berendsen, D.~van~der Spoel, and R.~van Drunen, ``Gromacs: a
  message-passing parallel molecular dynamics implementation,'' {\em Computer
  physics communications}, vol.~91, no.~1-3, pp.~43--56, 1995.

\bibitem{martinez2009packmol}
L.~Mart{\'\i}nez, R.~Andrade, E.~G. Birgin, and J.~M. Mart{\'\i}nez, ``Packmol:
  a package for building initial configurations for molecular dynamics
  simulations,'' {\em Journal of computational chemistry}, vol.~30, no.~13,
  pp.~2157--2164, 2009.

\bibitem{bennett1976efficient}
C.~H. Bennett, ``Efficient estimation of free energy differences from monte
  carlo data,'' {\em Journal of Computational Physics}, vol.~22, no.~2,
  pp.~245--268, 1976.

\bibitem{singh1984approach}
U.~C. Singh and P.~A. Kollman, ``An approach to computing electrostatic charges
  for molecules,'' {\em Journal of Computational Chemistry}, vol.~5, no.~2,
  pp.~129--145, 1984.

\bibitem{besler1990atomic}
B.~H. Besler, K.~M. Merz~Jr, and P.~A. Kollman, ``Atomic charges derived from
  semiempirical methods,'' {\em Journal of Computational Chemistry}, vol.~11,
  no.~4, pp.~431--439, 1990.

\bibitem{frisch2013gaussian}
M.~Frisch, G.~Trucks, H.~Schlegel, G.~Scuseria, M.~Robb, J.~Cheeseman,
  G.~Scalmani, V.~Barone, B.~Mennucci, G.~Petersson, {\em et~al.}, ``Gaussian
  09, revision b. 01,'' 2013.

\bibitem{potoff1998critical}
J.~J. Potoff and A.~Z. Panagiotopoulos, ``Critical point and phase behavior of
  the pure fluid and a lennard-jones mixture,'' {\em The Journal of chemical
  physics}, vol.~109, no.~24, pp.~10914--10920, 1998.

\bibitem{hartono2001prediction}
R.~Hartono, G.~A. Mansoori, and A.~Suwono, ``Prediction of solubility of
  biomolecules in supercritical solvents,'' {\em Chemical Engineering Science},
  vol.~56, no.~24, pp.~6949--6958, 2001.

\bibitem{de2009solid}
S.~V. de~Melo, G.~M.~N. Costa, A.~Viana, and F.~Pessoa, ``Solid pure component
  property effects on modeling upper crossover pressure for supercritical fluid
  process synthesis: A case study for the separation of annatto pigments using
  sc-co2,'' {\em The Journal of Supercritical Fluids}, vol.~49, no.~1,
  pp.~1--8, 2009.

\bibitem{su2006simulations}
Z.~Su and M.~Maroncelli, ``Simulations of solvation free energies and
  solubilities in supercritical solvents,'' {\em The Journal of chemical
  physics}, vol.~124, no.~16, p.~164506, 2006.

\bibitem{komkoua2013evaluation}
A.~Komkoua~Mbienda, C.~Tchawoua, D.~Vondou, and F.~Mkankam~Kamga, ``Evaluation
  of vapor pressure estimation methods for use in simulating the dynamic of
  atmospheric organic aerosols,'' {\em International Journal of Geophysics},
  vol.~2013, 2013.

\bibitem{o2014assessment}
S.~O'Meara, A.~M. Booth, M.~H. Barley, D.~Topping, and G.~McFiggans, ``An
  assessment of vapour pressure estimation methods,'' {\em Physical Chemistry
  Chemical Physics}, vol.~16, no.~36, pp.~19453--19469, 2014.

\bibitem{perlovich2004thermodynamics}
G.~L. Perlovich, S.~V. Kurkov, L.~K. Hansen, and A.~Bauer-Brandl,
  ``Thermodynamics of sublimation, crystal lattice energies, and crystal
  structures of racemates and enantiomers:(+)-and ($\pm$)-ibuprofen,'' {\em
  Journal of pharmaceutical sciences}, vol.~93, no.~3, pp.~654--666, 2004.

\bibitem{perlovich2004solvation}
G.~L. Perlovich, S.~V. Kurkov, A.~N. Kinchin, and A.~Bauer-Brandl, ``Solvation
  and hydration characteristics of ibuprofen and acetylsalicylic acid,'' {\em
  Aaps Pharmsci}, vol.~6, no.~1, pp.~22--30, 2004.

\bibitem{drozd2017novel}
K.~V. Drozd, A.~N. Manin, A.~V. Churakov, and G.~L. Perlovich, ``Novel
  drug--drug cocrystals of carbamazepine with para-aminosalicylic acid:
  Screening, crystal structures and comparative study of carbamazepine
  cocrystal formation thermodynamics,'' {\em CrystEngComm}, vol.~19, no.~30,
  pp.~4273--4286, 2017.

\bibitem{cao2008use}
X.~Cao, N.~Leyva, S.~R. Anderson, and B.~C. Hancock, ``Use of prediction
  methods to estimate true density of active pharmaceutical ingredients,'' {\em
  International journal of pharmaceutics}, vol.~355, no.~1-2, pp.~231--237,
  2008.

\bibitem{ardjmand2014measurement}
M.~Ardjmand, M.~Mirzajanzadeh, and F.~Zabihi, ``Measurement and correlation of
  solid drugs solubility in supercritical systems,'' {\em Chinese Journal of
  Chemical Engineering}, vol.~22, no.~5, pp.~549--558, 2014.

\bibitem{baum1997chemical}
E.~Baum, {\em Chemical property estimation: theory and application}.
\newblock CRC Press, 1997.

\bibitem{zielenkiewicz1999vapour}
X.~Zielenkiewicz, G.~Perlovich, and M.~Wszelaka-Rylik, ``The vapour pressure
  and the enthalpy of sublimation: determination by inert gas flow method,''
  {\em Journal of thermal analysis and calorimetry}, vol.~57, no.~1,
  pp.~225--234, 1999.

\bibitem{lyman1990handbook}
W.~J. Lyman, W.~F. Reehl, and D.~H. Rosenblatt, ``Handbook of chemical property
  estimation methods,'' 1990.

\bibitem{charoenchaitrakool2000micronization}
M.~Charoenchaitrakool, F.~Dehghani, N.~Foster, and H.~Chan, ``Micronization by
  rapid expansion of supercritical solutions to enhance the dissolution rates
  of poorly water-soluble pharmaceuticals,'' {\em Industrial \& engineering
  chemistry research}, vol.~39, no.~12, pp.~4794--4802, 2000.

\bibitem{huang2004solubility}
Z.~Huang, W.~D. Lu, S.~Kawi, and Y.~C. Chiew, ``Solubility of aspirin in
  supercritical carbon dioxide with and without acetone,'' {\em Journal of
  Chemical \& Engineering Data}, vol.~49, no.~5, pp.~1323--1327, 2004.

\bibitem{champeau2016solubility}
M.~Champeau, J.-M. Thomassin, C.~J\'{e}r\^{o}me, and T.~Tassaing, ``Solubility
  and speciation of ketoprofen and aspirin in supercritical co2 by infrared
  spectroscopy,'' {\em Journal of Chemical \& Engineering Data}, vol.~61,
  no.~2, pp.~968--978, 2016.

\bibitem{yamini2001solubilities}
Y.~Yamini, J.~Hassan, and S.~Haghgo, ``Solubilities of some nitrogen-containing
  drugs in supercritical carbon dioxide,'' {\em Journal of Chemical \&
  Engineering Data}, vol.~46, no.~2, pp.~451--455, 2001.

\bibitem{li2013new}
J.-h. Li, Z.~Huang, J.-l. Wei, and L.~Xu, ``A new optimization method for
  parameter determination in modeling solid solubility in supercritical co2,''
  {\em Fluid Phase Equilibria}, vol.~344, pp.~117--124, 2013.

\bibitem{klincewicz1984estimation}
K.~Klincewicz and R.~Reid, ``Estimation of critical properties with group
  contribution methods,'' {\em AIChE Journal}, vol.~30, no.~1, pp.~137--142,
  1984.

\bibitem{joback1987estimation}
K.~G. Joback and R.~C. Reid, ``Estimation of pure-component properties from
  group-contributions,'' {\em Chemical Engineering Communications}, vol.~57,
  no.~1-6, pp.~233--243, 1987.

\bibitem{constantinou1994new}
L.~Constantinou and R.~Gani, ``New group contribution method for estimating
  properties of pure compounds,'' {\em AIChE Journal}, vol.~40, no.~10,
  pp.~1697--1710, 1994.

\bibitem{jahromi2019estimation}
S.~A. Jahromi and A.~Roosta, ``Estimation of critical point, vapor pressure and
  heat of sublimation of pharmaceuticals and their solubility in supercritical
  carbon dioxide,'' {\em Fluid Phase Equilibria}, vol.~488, pp.~1--8, 2019.

\bibitem{kontogeorgis1997method}
G.~M. Kontogeorgis, I.~Smirlis, I.~V. Yakoumis, V.~Harismiadis, and D.~P.
  Tassios, ``Method for estimating critical properties of heavy compounds
  suitable for cubic equations of state and its application to the prediction
  of vapor pressures,'' {\em Industrial \& engineering chemistry research},
  vol.~36, no.~9, pp.~4008--4012, 1997.

\bibitem{panagiotopoulos1988phase}
A.~Z. Panagiotopoulos, N.~Quirke, M.~Stapleton, and D.~Tildesley, ``Phase
  equilibria by simulation in the gibbs ensemble: alternative derivation,
  generalization and application to mixture and membrane equilibria,'' {\em
  Molecular Physics}, vol.~63, no.~4, pp.~527--545, 1988.

\bibitem{merker2008comment}
T.~Merker, J.~Vrabec, and H.~Hasse, ``Comment on “an optimized potential for
  carbon dioxide”[j. chem. phys. 122, 214507 (2005)],'' {\em The Journal of
  chemical physics}, vol.~129, no.~8, p.~214507, 2008.

\bibitem{fedorova2016conformational}
I.~Fedorova, D.~Ivlev, and M.~Kiselev, ``Conformational lability of ibuprofen
  in supercritical carbon dioxide,'' {\em Russian Journal of Physical Chemistry
  B}, vol.~10, no.~7, pp.~1153--1162, 2016.

\bibitem{schmid2011definition}
N.~Schmid, A.~P. Eichenberger, A.~Choutko, S.~Riniker, M.~Winger, A.~E. Mark,
  and W.~F. van Gunsteren, ``Definition and testing of the gromos force-field
  versions 54a7 and 54b7,'' {\em European biophysics journal}, vol.~40, no.~7,
  pp.~843--856, 2011.

\bibitem{andersen1971relationship}
H.~C. Andersen, J.~D. Weeks, and D.~Chandler, ``Relationship between the
  hard-sphere fluid and fluids with realistic repulsive forces,'' {\em Physical
  Review A}, vol.~4, no.~4, p.~1597, 1971.

\end{thebibliography}

\clearpage

\end{document}